\documentclass[11pt]{article}
\usepackage{amsmath}
\usepackage{acl}
\usepackage{times}
\usepackage{latexsym}
\usepackage{float}
\usepackage{multicol}
\usepackage[T1]{fontenc}
\usepackage[utf8]{inputenc}
\usepackage{stfloats}
\usepackage{microtype}

\usepackage{inconsolata}
\usepackage{xcolor}
\usepackage{tcolorbox}
\tcbuselibrary{breakable}

\usepackage{booktabs} % 必选，用于专业表格
\usepackage{pgfplots} % 选选，用于绘制 Heatmap
\usepackage{multirow}
\usepackage{placeins}

\usepackage{makecell}       
\usepackage[table]{xcolor} 
\usepackage{longtable}
\usepackage{enumitem}
\usepackage{array} 

\usepackage{amsmath}
\usepackage{dsfont}
\usepackage{amssymb}

\usepackage{listings}
\lstdefinestyle{pycode}{
  language=Python,
  basicstyle=\ttfamily\small,
  keywordstyle=\color{blue!60!black},
  commentstyle=\color{gray!70!black},
  stringstyle=\color{teal!60!black},
  numbers=left,
  numberstyle=\tiny\color{gray!70!black},
  stepnumber=1,
  numbersep=6pt,
  showstringspaces=false,
  breaklines=true,
  breakatwhitespace=true,
  frame=single,
  rulecolor=\color{gray!35},
  tabsize=2,
  captionpos=b
}

\pgfplotsset{compat=1.18}
\usepackage{hyperref}
\title{RSA-Bench: Benchmarking Audio Large Models in Real-World Acoustic Scenarios}

\author{
    \textbf{Yibo Zhang}$^{1,}$\thanks{ \ \ Equal contribution.}, 
    \textbf{Liang Lin}$^{2,*}$, 
    \textbf{Kaiwen Luo}$^{2,*}$, 
    \textbf{Shilinlu Yan}$^{1}$, 
    \textbf{Jin Wang}$^{3}$, 
    \textbf{Yaoqi Guo}$^{2}$,\\ 
    \textbf{Yitian Chen}$^{4}$, 
    \textbf{Yalan Qin}$^{4}$, 
    \textbf{Zhenhong Zhou}$^{2}$, 
    \textbf{Kun Wang}$^{2}$,
    \textbf{Li Sun}$^{1,}$\thanks{ \ \ Corresponding author.}
    \vspace{0.2cm} \\
    \textsuperscript{1}Beijing University of Posts and Telecommunications,\\
    \textsuperscript{2}Nanyang Technological University,\\
    \textsuperscript{3}Xidian University, 
    \textsuperscript{4}Shanghai University\\[0.5ex]
    \textbf{Correspondence:} \texttt{zhangyibo2023@bupt.edu.cn}
}

\definecolor{ASRcolor}{HTML}{D45959}
\definecolor{GRcolor}{HTML}{2F74B8}
\definecolor{ERcolor}{HTML}{F9B43F}
\definecolor{MRcolor}{HTML}{897CD3}
\definecolor{SQAcolor}{HTML}{D2B48C}
\definecolor{SIcolor}{HTML}{5DCE9C}

\newcommand{\hlToken}[1]{\textcolor{magenta}{\textbf{#1}}}

\begin{document}
\maketitle

\begin{abstract}
While Audio Large Models (ALLMs) have achieved remarkable proficiency, their robustness remains brittle in real-world deployment. 
Existing evaluations largely rely on synthetic Gaussian noise or simplistic single-source interference, failing to capture the intricate, multi-layered acoustic dynamics---or ``Acoustic Ecology''---that characterize authentic physical environments. 
To bridge this ecological gap, we introduce \textbf{RSA-Bench}, a comprehensive robustness benchmark designed to stress-test ALLMs through high-fidelity auditory scene simulations. 
Unlike traditional methods, we construct evaluation samples by naturally superimposing diverse environmental soundscapes---spanning \textit{Pasture}, \textit{Extreme Weather}, \textit{Classroom}, and \textit{Outdoors}---onto clean speech signals across a spectrum of interference intensities. 
By evaluating models on six core tasks ranging from fundamental perception to complex reasoning, our study unveils three macro-level insights: 
\textbf{(I) The Perception-Cognition Gap:} Models maintain relative resilience in low-level recognition but suffer a \textbf{functional collapse} in high-order reasoning tasks under stress; 
\textbf{(II) Scenario Sensitivity:} ``Vocal-like'' interference (e.g., children playing) proves significantly more destructive than mechanical noise, challenging the model's auditory attention mechanisms; 
and \textbf{(III) The Denoising Paradox:} Standard speech enhancement often exacerbates performance degradation, as ALLMs prove highly sensitive to the semantic distortions introduced by denoising artifacts.
Our code and dataset are publicly available at \url{https://github.com/Yibo124/RSA-Bench}.

\end{abstract}

\section{Introduction}

\begin{figure*}[t] % 注意这里加了星号 *
    \centering
    % width=\linewidth 或 width=\textwidth 在 figure* 环境中都会指代整个页面的宽度
    \includegraphics[width=\linewidth]{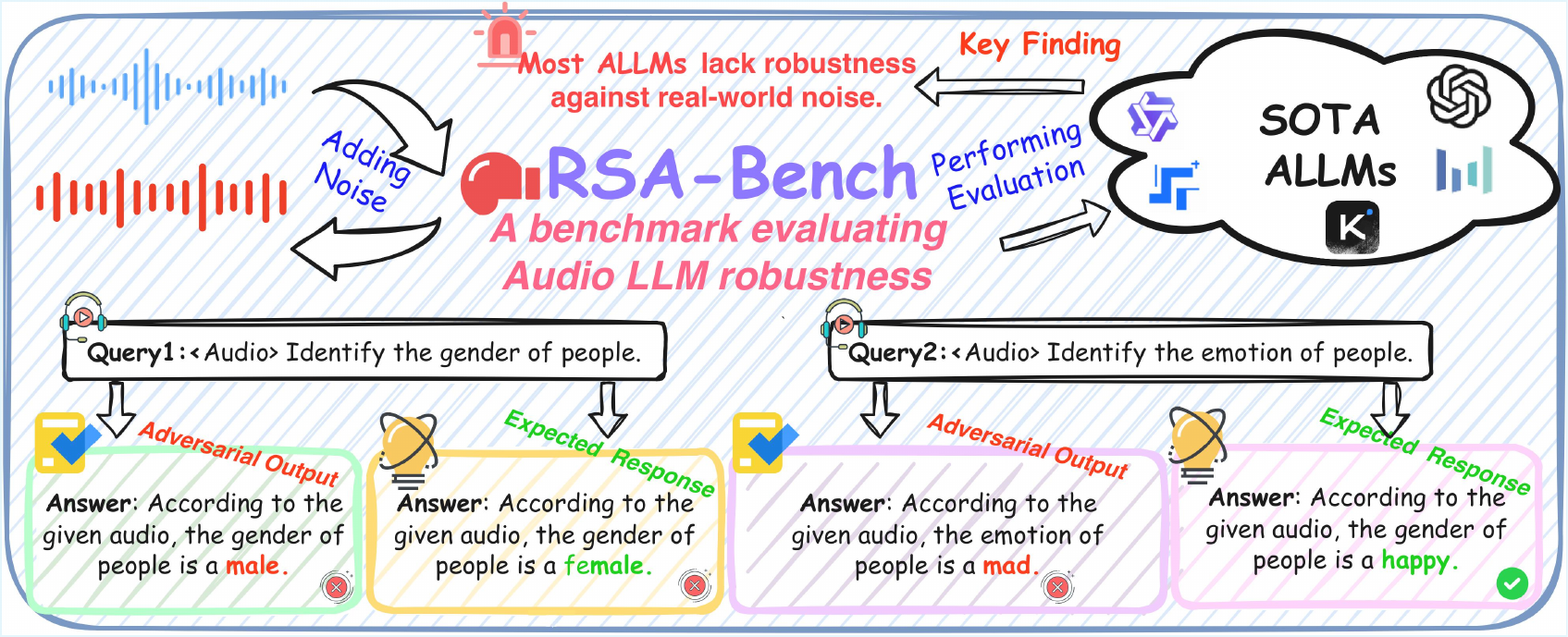}
    \vspace{-1.6em}
    \caption{A framework of our \textbf{RSA-Benchmark} for evaluating Audio-LLM robustness across six different tasks. }
    \label{fig:introduction}
    \vspace{-1em}
\end{figure*} % 结束也要加星号

In recent years, the intersection of Large Language Models (LLMs) and audio processing has given rise to Audio Large Models (ALLMs) \citep{goel2025audio,yang2025audio,yang2025towards}. By integrating audio encoders with pre-trained LLMs, these models have demonstrated remarkable capabilities across a wide range of tasks, including Automatic Speech Recognition (ASR) \citep{ahlawat2025automatic,fatehifar2025applications,liu2025systematic} , speech translation \citep{sarim2025direct}, and audio-based reasoning \citep{xie2025audio}. Cutting-edge models have achieved impressive performance on standard benchmarks \citep{wang2025audiobench,kumar2025mmau,ma2025mmar}, exhibiting strong semantic understanding and instruction-following abilities when processing high-quality audio inputs.

However, the promising results obtained in controlled, noise-free environments often fail to translate to real-world deployment scenarios \citep{wang2025advancing,atwany2025lost}. Real-world acoustic environments are characterized by diverse, unavoidable background noises and multi-source interference. While previous works have established benchmarks for general audio capabilities \citep{yang2024air,wang2025audiobench,ahia2025blab}, there is a systematic absence of evaluations that quantify how ALLMs behave under acoustic stress. Existing resources fail to reflect the complexity of a true "Acoustic Ecology," \citep{wrightson2000introduction,pace2025acoustic} where target signals are inextricably intertwined with diverse background sounds. \textbf{Specifically, the magnitude of the performance gap between ideal and noisy conditions in ALLMs} has not been sufficiently quantified, leaving the true robustness of these models in question. 

To address this fundamental limitation, we present \textit{RSA-Bench}, a  robustness benchmark designed to stress-test ALLMs within complex acoustic scenarios. Distinguished by its scale, the dataset covers more than 100,000 samples across six core tasks, ranging from basic ASR to high-order reasoning such as Math and QA. Specifically, the benchmark features a high-fidelity ``Acoustic Ecology'' constructed from four distinct environments: Pasture, Extreme Weather, Classroom, and Outdoor. To ensure realism, a multi-source superposition strategy is employed, naturally mixing 1 to 4 noise sources with the original signal. Ultimately, this setting prioritizes ecological validity over artificial difficulty, simulating the complexity of real-world environments where target signals are inextricably intertwined with diverse background sounds.

As shown in Figure \ref{fig:introduction}, our study reveals a stark contrast in model performance between clean and noisy inputs. We observe that most models exhibit a precipitous decline in capabilities as the acoustic environment becomes more complex, exposing a widespread vulnerability across current architectures. The degradation is particularly severe in tasks requiring precise semantic reasoning. Regarding mitigation, we applied four standard denoising methods to the noisy audio in an attempt to alleviate the negative impact of interference. However, we find that real-world noise proves to be remarkably persistent. Standard methods often struggle to effectively strip away this interference; instead, the attempt may disrupt the semantic integrity of the original audio, potentially leading to performance that is not only unrestored but further degraded. 

\noindent \textbf{Experimental Takeaways.} 
\begin{itemize}[leftmargin=*, noitemsep, topsep=0pt, parsep=0pt]
    \item \textbf{Widespread Robustness Vulnerability.} 
     \textit{RSA-benchmark} reveals a universal performance decline across diverse interference types, confirming that high capabilities in clean environments fail to translate to reliability in complex physical-world deployment.

    \item \textbf{The Perception-Cognition Gap.} 
    Acoustic interference disproportionately impacts cognitive over perceptual capabilities. While models retain resilience in low-level tasks like gender recognition, they suffer a \textbf{functional collapse} in high-order reasoning under stress, exposing a critical bottleneck in complex semantic processing.

    \item \textbf{The Denoising Paradox.} 
    External mitigation strategies often prove counterproductive. We find that standard speech enhancement algorithms frequently \textbf{exacerbate} errors, as ALLMs are significantly more sensitive to the spectral artifacts introduced by denoising than to the natural background noise itself.
\end{itemize}

\section{Related Work}

\textbf{Audio Large Models.}
The landscape of audio processing has shifted dramatically from specialized models to general-purpose ALLMs \citep{zhang2023speechgpt,chu2024qwen2,huang2024audiogpt}. Early works primarily focused on discriminative tasks such as ASR \citep{ahlawat2025automatic,he2025survey}. Recently, the integration of audio encoders with LLMs has empowered models like GPT-4o-Audio \citep{hurst2024gpt} and Qwen2-Audio \citep{chu2024qwen2} to perform reasoning, instruction following, and multi-turn dialogue. The emergence of models like Qwen2.5-Omni \citep{xu2025qwen2}further exemplifies the trend towards unified multimodal understanding \citep{zhang2025deepaudio}, where models process audio, text, and other modalities within a single end-to-end framework.

\textbf{Robustness against Acoustic Interference.}
Robustness has been a longstanding pursuit in signal processing, traditionally measured by Word Error Rate (WER) in ASR systems under low Signal-to-Noise Ratio (SNR) \citep{song2025nonlinear,akomodi2025statistical} conditions. In the era of ALLMs, the scope of robustness extends beyond recognition accuracy to encompass comprehensive understanding and reasoning capabilities in noisy contexts. Recent empirical studies have begun to explore the negative impact of audio interference. For instance, recent work demonstrated that environmental noise can be utilized to bypass model safety mechanisms \citep{zhang2025enj,peng2025jalmbench,chen2025audiojailbreak}, while other studies investigated how irrelevant audio acts as a distractor for text-based reasoning \citep{li2025silence}. However, the systematic impact of environmental noise on audio-centric cognitive tasks remains under-explored \citep{yang2024air,wang2025audiobench}. Furthermore, while speech enhancement (SE) \citep{yousif2025speech,jannu2025overview,huang2025advances}is a solution in traditional pipelines, its interaction with large-scale pre-trained encoders is complex. Our work provides a quantitative gap analysis and empirically examines the effectiveness of denoising methods. We find that applying off-the-shelf enhancement tools often fails to recover performance \citep{chondhekar2025noising}, highlighting both the stubborn persistence of acoustic interference and the sensitivity of ALLMs to semantic distortions introduced by enhancement artifacts.

\section{RSA-Bench}

% \subsection{Overall}
% \label{sec:overall}

%Existing benchmarks primarily evaluate ALLMs using clean, high-quality audio recordings, focusing on expanding task breadth or context length under ideal conditions. However, this approach fails to capture the unpredictability of real-world acoustic environments, leaving the model's robustness in practical deployment largely unverified. 
Uniquely, \textit{RSA-Bench} establishes the first framework to systematically investigate ALLM robustness against complex environmental noise. Instead of theoretical simulations, we construct four distinct real-world acoustic scenarios, each composed of representative audio elements designed to challenge specific aspects of model stability:

\begin{itemize}[leftmargin=*, noitemsep, topsep=0pt, parsep=0pt]
    \item \textbf{Pasture:} Represents an environment dominated by irregular animal vocalizations. We explicitly select non-stationary sounds from \textbf{cows}, \textbf{dogs}, \textbf{hens}, and \textbf{sheep} to test the model's stability against sudden biological sounds.
    
    \item \textbf{Extreme Weather:} Simulates a complex acoustic environment with mixed interference types. This scenario combines continuous \textbf{heavy rain} and \textbf{wind} with sudden \textbf{thunderstorms} and tonal \textbf{wind chimes}, evaluating the model's stability under varying acoustic pressure.
    
    \item \textbf{Classroom:} Replicates an indoor environment characterized by subtle but persistent human activity. We incorporate rhythmic \textbf{clock ticking} alongside sporadic human-generated noises such as \textbf{coughing}, \textbf{keyboard typing}, and \textbf{drinking}, simulating a scenario where background activities compete with the target speech.
    
    \item \textbf{Outdoors:} Represents an open-air environment. To ensure ecological fidelity, we synthesize a soundscape featuring \textbf{children playing}, \textbf{bird chirping}, \textbf{flowing streams}, and texture-specific \textbf{footsteps on grass}. This tests the model's adaptability to unstructured acoustic events.
\end{itemize}

\begin{figure}[t!]
    \centering
    % \linewidth 在 figure 环境中指的是当前一栏的宽度
    \includegraphics[width=\linewidth]{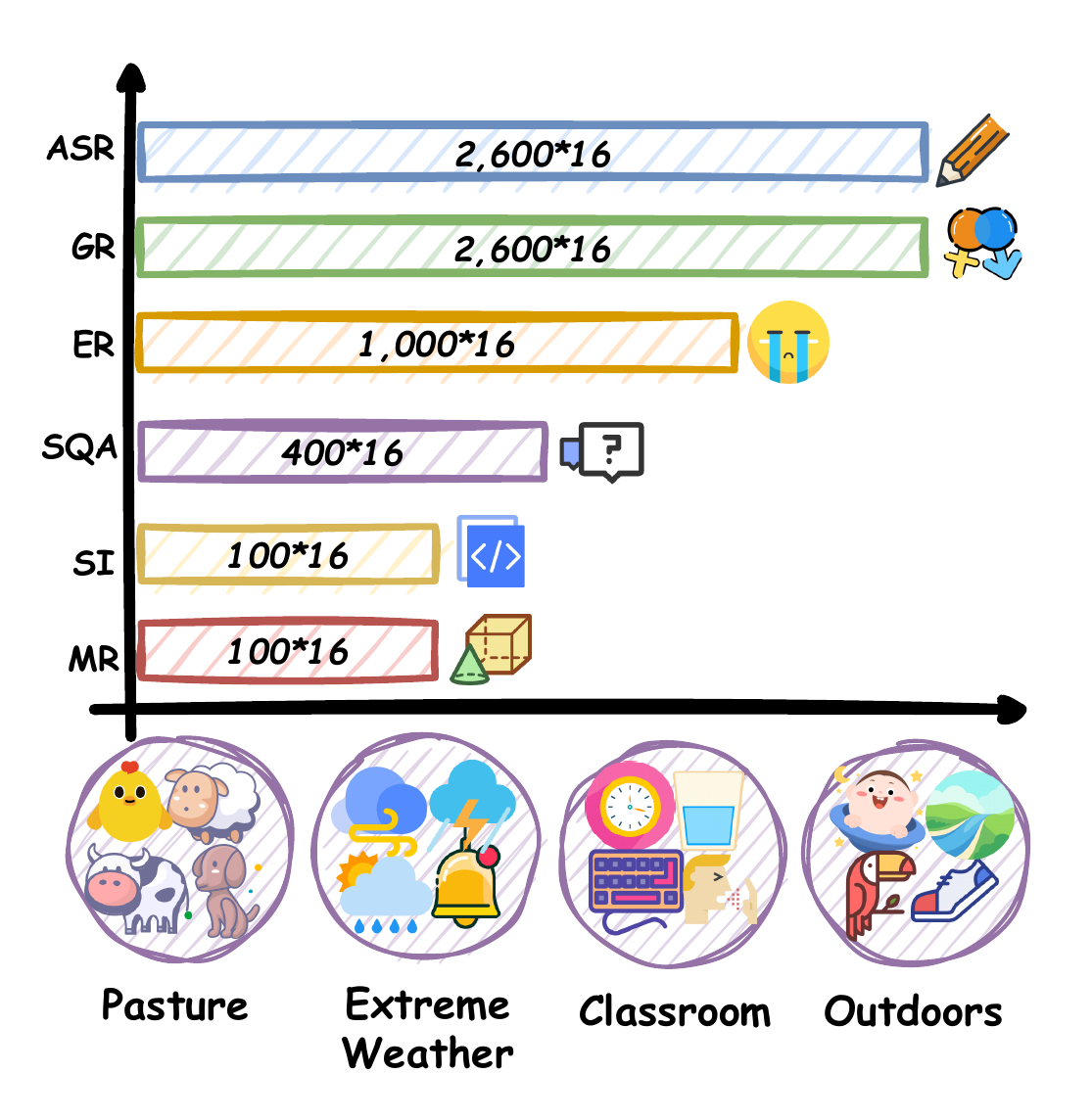}
    \vspace{-1.7em}
    \caption{Overview of the RSA-Bench data composition, which covers 6 tasks, 4 real-world acoustic scenarios, and totals over 100,000 samples.}
    \label{fig:task}
    \vspace{-1.5em}
\end{figure}

\subsection{Data Construction}
\label{sec:data_construction}

To systematically evaluate model robustness, we construct four distinct variations for each of the four predefined acoustic scenarios by varying the number of superimposed real-world interference, specifically setting $K\in\{1, 2, 3, 4\}$ to represent increasing levels of environmental complexity. For each individual audio sample in the dataset, this construction process follows four sequential steps: source collection, temporal alignment, energy alignment, and superposition.

\paragraph{Step 1. Source Collection.}
The construction of RSA-Bench begins with the curation of high-quality source materials. We aggregate data from two distinct sources:
\begin{itemize}[leftmargin=*, noitemsep, topsep=0pt, parsep=0pt]
    \item \textbf{Clean Audio Stream:} We select samples from six representative datasets covering both perception tasks and reasoning tasks.
    \item \textbf{Noise Audio Stream:} To simulate authentic acoustic ecologies, we utilize recordings from the Environmental Sound Classification (ESC) \citep{piczak2015esc} subset of DynamicSuperb. These are manually categorized into four distinct scenarios: \textit{Pasture}, \textit{Extreme Weather}, \textit{Classroom}, and \textit{Outdoor}.
    %噪声数据集这里注意！！能不能这么写
    
\end{itemize}

\paragraph{Step 2. Temporal Alignment.}
Upon obtaining the source materials, we define the discrete-time clean audio signal $s[n]$ of length $N$. For each sample, we randomly select $K$ noise clips from a target environmental category ($K \in \{1, \dots, 4\}$). Let $w_k[n]$ represent the $k$-th raw noise signal of length $M_k$. To address the duration mismatch between the clean audio and the noise, we apply a temporal alignment operator. We generate the aligned noise sequence $\tilde{w}_k[n]$ using a modulo operation:
\begin{equation}
    \tilde{w}_k[n] = w_k[n \bmod M_k], \quad \text{for } 0 \le n < N.
\end{equation}
This formulation unifies two behaviors: if the noise is shorter than the clean audio ($M_k < N$), it is cyclically tiled to fill the duration; if the noise is longer ($M_k > N$), it is automatically truncated to match the target length $N$. This ensures continuous background coverage.

\paragraph{Step 3. RMS-based Energy Alignment.}
To establish a consistent interference intensity, we normalize the energy of the noise audio to strictly match that of the clean audio. We first calculate the Root Mean Square (RMS) energy for the clean audio ($R_s$) and the aligned noise audio ($R_{w_k}$):
\begin{equation}
    R_s = \sqrt{\frac{1}{N}\sum_{n=0}^{N-1} s^2[n]}, \quad R_{w_k} = \sqrt{\frac{1}{N}\sum_{n=0}^{N-1} \tilde{w}_k^2[n]}.
\end{equation}
We then derive an adaptive scaling factor $\lambda_k$ to align the noise energy to the speech energy:
\begin{equation}
    \lambda_k = \frac{R_s}{R_{w_k}}.
\end{equation}
%参数要明确点出来，这个加在哪里合适一点
In our experiments, we fix $\lambda_k = 1$. This setup yields an SNR range comparable with major benchmarks like WHAM! , WHAMR! , and LibriMix, \citep{wichern2019wham,maciejewski2020whamr,cosentino2020librimix} ensuring a standard and rigorous evaluation of environmental robustness.

\paragraph{Step 4. Superposition and Dynamic Constraint.}
Finally, we generate the noisy speech sample by linearly superimposing the clean audio and the scaled real-world interference. To ensure the audio data remains within the valid amplitude range, we apply a hard constraint function. The final evaluation sample $x[n]$ is formulated as:
\begin{equation}
    x[n] = \operatorname{clip}\left( s[n] + \sum_{k=1}^{K} \left( \tilde{w}_k[n] \cdot \lambda_k \right), \;-1, \;1 \right),
\end{equation}
where $\operatorname{clip}(v, -1, 1)$ restricts the amplitude values to the interval $[-1, 1]$.

\vspace{0.5em}

\noindent By executing the aforementioned four steps for each clean sample across all scenarios and the four interference levels ($K=1$ to $4$), this combinatorial design results in $4 \text{ scenarios} \times 4 \text{ intensity levels} = 16$ unique configurations per original sample. Together with the original clean version, RSA-Bench provides a total of 17 test conditions per sample, enabling a fine-grained analysis of model robustness as environmental complexity scales.

\subsection{Task Taxonomy and Definitions}
\label{sec:tasks}

To comprehensively disentangle the impact of environmental complexity on different model capabilities, we categorize the six evaluation tasks into two distinct \textbf{categories}: \textbf{Perception \& Paralinguistics} and \textbf{Cognitive Reasoning}. 

\subsubsection{Perception \& Paralinguistics.}
These tasks assess the model's fundamental ability to perceive acoustic signals and extract specific attributes, evaluating whether the model can maintain signal fidelity under environmental interference.

\noindent\textbf{ASR.} ASR aims to transcribe spoken content into verbatim text. This task measures the model's robustness in preserving linguistic information against environmental masking. We use source samples from LibriSpeech \citep{panayotov2015librispeech} to evaluate phonetic recognition accuracy under complex acoustic conditions.

\noindent\textbf{Gender Recognition (GR).} This task evaluates the ability to discern speaker identity traits based on vocal characteristics. Established upon the IEMOCAP dataset \citep{busso2008iemocap}, it challenges the model to isolate the speaker's biological features from background environments, testing the robustness of acoustic feature extraction.

\noindent\textbf{Emotion Recognition (ER).} Emotion is a critical paralinguistic element conveyed through prosody and tone. Utilizing MELD \citep{poria2019meld} as the source, this task requires the model to interpret the speaker's emotional state. 
\begin{table*}[t]
\centering
\captionsetup{font=scriptsize, skip=1pt}
\scriptsize
\setlength{\tabcolsep}{2.4pt}
\renewcommand{\arraystretch}{0.95} % 调至 0.95 达到紧凑效果

% 核心修复：消除booktabs带来的白边缝隙
\setlength{\aboverulesep}{0pt}
\setlength{\belowrulesep}{0pt}
\setlength{\cmidrulekern}{0.2em}

% 定义颜色
\definecolor{head_light}{HTML}{F5F5F5} % 标题浅灰
\definecolor{color_asr}{HTML}{E6F0FF}  % 蓝
\definecolor{color_er}{HTML}{E6FFEF}   % 绿
\definecolor{color_gr}{HTML}{FFF3E6}   % 橙
\definecolor{color_mr}{HTML}{F2E6FF}   % 紫
\definecolor{color_sqa}{HTML}{FFE6EF}  % 粉
\definecolor{color_si}{HTML}{F0FFE6}   % 浅绿

\newcommand{\ua}[1]{\textcolor{red}{\tiny{$\uparrow_{#1}$}}}
\newcommand{\da}[1]{\textcolor{green!65!black}{\tiny{$\downarrow_{#1}$}}}

\resizebox{\textwidth}{!}{%
\begin{tabular}{l *{11}{c}}
\hline
\rowcolor{head_light}
\textbf{$K$} & \multicolumn{11}{c}{\textbf{Models}} \\ 
\rowcolor{head_light} \cmidrule(lr){2-12}
\rowcolor{head_light}
& \textbf{Qwen2-Audio} & \textbf{SALMONN} & \textbf{SeaLLMs} & \textbf{Phi-4} & \textbf{MERaLION} & \textbf{StepAudio2} & \textbf{MiniCPM} & \textbf{Qwen-Turbo} & \textbf{Qwen2.5-Omni} & \textbf{Qwen3-Omni} & \textbf{GPT-4o-Audio} \\
\hline

% ASR 部分
\rowcolor{color_asr} \multicolumn{12}{l}{\textit{ASR (WER $\downarrow$)}} \\ \hline
\rowcolor{color_asr} \textit{$K=0$} & 3.45 & 10.49 & 5.52 & \textbf{1.67} & 2.34 & 3.90 & 2.95 & 23.78 & 23.32 & 1.72 & \underline{50.01} \\
\rowcolor{color_asr} \textit{$K=1$} & 8.49 \ua{5.04} & 24.47 \ua{13.98} & 25.49 \ua{19.97} & 7.07 \ua{5.40} & 11.63 \ua{9.29} & 7.59 \ua{3.69} & 21.08 \ua{18.13} & 27.95 \ua{4.17} & 28.89 \ua{5.57} & \textbf{5.70} \ua{3.98} & \underline{64.69} \ua{14.68} \\
\rowcolor{color_asr} \textit{$K=2$} & 19.67 \ua{16.22} & \underline{124.79} \ua{114.3} & 51.13 \ua{45.61} & \textbf{19.54} \ua{17.87} & 30.89 \ua{28.55} & 20.47 \ua{16.57} & 57.34 \ua{54.39} & 42.30 \ua{18.52} & 45.18 \ua{21.86} & 48.41 \ua{46.69} & 86.97 \ua{36.96} \\
\rowcolor{color_asr} \textit{$K=3$} & 35.97 \ua{32.52} & \underline{317.33} \ua{306.8} & 125.50 \ua{119.9} & 42.89 \ua{41.22} & 55.35 \ua{53.01} & \textbf{34.49} \ua{30.59} & 89.09 \ua{86.14} & 61.54 \ua{37.76} & 61.57 \ua{38.25} & 259.56 \ua{257.8} & 107.27 \ua{57.26} \\
\rowcolor{color_asr} \textit{$K=4$} & \textbf{54.75} \ua{51.30} & 509.11 \ua{498.6} & 279.27 \ua{273.7} & 81.12 \ua{79.45} & 76.04 \ua{73.70} & 66.67 \ua{62.77} & 121.17 \ua{118.2} & 96.42 \ua{72.64} & 93.27 \ua{69.95} & \underline{557.20} \ua{555.5} & 118.39 \ua{68.38} \\
\hline

% ER 部分
\rowcolor{color_er} \multicolumn{12}{l}{\textit{ER (Score $\uparrow$)}} \\ \hline
\rowcolor{color_er} \textit{$K=0$} & 51.53 & 40.53 & 47.80 & 49.92 & 52.60 & \textbf{56.81} & 55.32 & 52.99 & 52.91 & 47.20 & \underline{30.61} \\
\rowcolor{color_er} \textit{$K=1$} & 35.29 \da{16.24} & 30.87 \da{9.66} & 20.91 \da{26.89} & 22.79 \da{27.13} & \textbf{52.56} \da{0.04} & 38.69 \da{18.12} & 30.03 \da{25.29} & 22.91 \da{30.08} & 23.18 \da{29.73} & 34.10 \da{13.10} & \underline{5.67} \da{24.94} \\
\rowcolor{color_er} \textit{$K=2$} & 35.24 \da{16.29} & 30.45 \da{10.08} & 20.37 \da{25.43} & 23.86 \da{26.06} & \textbf{55.63} \ua{3.03} & 38.54 \da{18.27} & 29.34 \da{25.98} & 25.02 \da{27.97} & 24.90 \da{28.01} & 33.56 \da{13.64} & \underline{8.93} \da{21.68} \\
\rowcolor{color_er} \textit{$K=3$} & 30.57 \da{20.96} & 30.45 \da{10.08} & 15.63 \da{32.17} & 18.16 \da{31.76} & \textbf{46.51} \da{6.09} & 36.74 \da{20.07} & 29.84 \da{25.48} & 14.10 \da{38.89} & 13.60 \da{39.31} & 33.41 \da{13.79} & \underline{1.07} \da{29.54} \\
\rowcolor{color_er} \textit{$K=4$} & 29.46 \da{22.07} & 30.22 \da{10.31} & 15.05 \da{32.75} & 13.90 \da{36.02} & \textbf{45.40} \da{7.20} & 36.78 \da{20.03} & 29.42 \da{25.90} & 10.57 \da{42.42} & 12.34 \da{40.57} & 34.10 \da{13.10} & \underline{0.23} \da{30.38} \\
\hline

% GR 部分
\rowcolor{color_gr} \multicolumn{12}{l}{\textit{GR (Score $\uparrow$)}} \\ \hline
\rowcolor{color_gr} \textit{$K=0$} & \textbf{96.02} & 82.37 & 79.87 & \underline{38.65} & 85.26 & 86.95 & 93.43 & 91.63 & 91.53 & 95.92 & \textcolor{gray}{--} \\
\rowcolor{color_gr} \textit{$K=1$} & \textbf{93.63} \da{2.40} & 82.67 \ua{0.30} & 71.04 \da{8.83} & \underline{38.65} \textcolor{gray}{--} & 82.97 \da{2.29} & 85.96 \da{0.99} & 91.33 \da{2.10} & 91.53 \da{0.10} & 91.04 \da{0.49} & 92.53 \da{3.39} & \textcolor{gray}{--} \\
\rowcolor{color_gr} \textit{$K=2$} & 90.04 \da{5.99} & 71.41 \da{10.96} & 74.19 \da{5.68} & \underline{32.07} \da{6.58} & 84.06 \da{1.20} & 81.67 \da{5.28} & 90.14 \da{3.29} & 89.84 \da{1.79} & 90.14 \da{1.39} & \textbf{91.33} \da{4.59} & \textcolor{gray}{--} \\
\rowcolor{color_gr} \textit{$K=3$} & 87.65 \da{8.38} & 65.84 \da{16.54} & 73.23 \da{6.64} & \underline{27.69} \da{10.96} & 81.77 \da{3.49} & 83.76 \da{3.19} & 85.16 \da{8.27} & 88.55 \da{3.08} & \textbf{89.54} \da{1.99} & 88.35 \da{7.57} & \textcolor{gray}{--} \\
\rowcolor{color_gr} \textit{$K=4$} & 81.77 \da{14.25} & 59.66 \da{22.71} & 75.66 \da{4.21} & \underline{21.41} \da{17.24} & 76.49 \da{8.77} & 77.19 \da{9.76} & 81.87 \da{11.56} & 86.35 \da{5.28} & 87.05 \da{4.48} & \textbf{88.94} \da{6.98} & \textcolor{gray}{--} \\

\hline

% MR 部分
\rowcolor{color_mr} \multicolumn{12}{l}{\textit{MR (Acc $\uparrow$)}} \\ \hline
\rowcolor{color_mr} \textit{$K=0$} & 66.00 & \underline{18.00} & 62.00 & 3.00 & 74.00 & 75.00 & 75.00 & 88.00 & 89.00 & 91.00 & \textbf{93.00} \\
\rowcolor{color_mr} \textit{$K=1$} & 43.00 \da{23.00} & \underline{5.00} \da{13.00} & 29.00 \da{33.00} & 1.00 \da{2.00} & 46.00 \da{28.00} & 47.00 \da{28.00} & 42.00 \da{33.00} & 48.00 \da{40.00} & 39.00 \da{50.00} & \textbf{63.00} \da{28.00} & 49.00 \da{44.00} \\
\rowcolor{color_mr} \textit{$K=2$} & 23.00 \da{43.00} & \underline{2.00} \da{16.00} & 10.00 \da{52.00.00} & 2.00 \da{1.00} & 27.00 \da{47.00} & 26.00 \da{49.00} & 18.00 \da{57.00} & 18.00 \da{70.00} & 16.00 \da{73.00} & \textbf{40.00} \da{51.00} & 16.00 \da{77.00} \\
\rowcolor{color_mr} \textit{$K=3$} & 6.00 \da{60.00} & \underline{0.00} \da{18.00} & 1.00 \da{61.00} & 2.00 \da{1.00} & 7.00 \da{67.00} & 12.00 \da{63.00} & 7.00 \da{68.00} & 7.00 \da{81.00} & 4.00 \da{85.00} & \textbf{18.00} \da{73.00} & 6.00 \da{8.007} \\
\rowcolor{color_mr} \textit{$K=4$} & 4.00 \da{62.00} & \underline{0.00} \da{18.00} & \underline{0.00} \da{62.00} & 1.00 \da{2.00} & 3.00 \da{71.00} & \textbf{6.00} \da{69.00} & 1.00 \da{74.00} & 4.00 \da{84.00} & 4.00 \da{85.00} & 5.00 \da{86.00} & 3.00 \da{90.00} \\
\hline

% SQA 部分
\rowcolor{color_sqa} \multicolumn{12}{l}{\textit{SQA (Score $\uparrow$)}} \\ \hline
\rowcolor{color_sqa} \textit{$K=0$} & 79.85 & 79.90 & \underline{78.58} & 85.74 & 80.69 & 81.37 & 82.94 & 82.50 & 83.82 & 80.98 & \textbf{86.62} \\
\rowcolor{color_sqa} \textit{$K=1$} & 77.21 \da{2.64} & \underline{73.14} \da{6.76} & 73.92 \da{4.66} & \textbf{86.42} \ua{0.68} & 80.29 \da{0.40} & 79.31 \da{2.06} & 82.45 \da{0.49} & 82.65 \ua{0.15} & 81.37 \da{2.45} & 80.83 \da{0.15} & 86.18 \da{0.44} \\
\rowcolor{color_sqa} \textit{$K=2$} & 73.97 \da{5.88} & 67.84 \da{12.06} & \underline{65.93} \da{12.65} & 80.69 \da{5.05} & 77.60 \da{3.09} & 76.47 \da{4.90} & 80.00 \da{2.94} & 79.71 \da{2.79} & 78.97 \da{4.85} & 81.03 \ua{0.05} & \textbf{86.18} \da{0.44} \\
\rowcolor{color_sqa} \textit{$K=3$} & 67.21 \da{12.64} & 63.82 \da{16.08} & \underline{58.38} \da{20.20} & 80.05 \da{5.69} & 73.68 \da{7.01} & 69.07 \da{12.30} & 75.78 \da{7.16} & 70.20 \da{12.30} & 73.53 \da{10.29} & 75.74 \da{5.24} & \textbf{83.38} \da{3.24} \\
\rowcolor{color_sqa} \textit{$K=4$} & 62.25 \da{17.60} & 62.75 \da{17.15} & \underline{55.74} \da{22.84} & 73.28 \da{12.46} & 71.08 \da{9.61} & 61.27 \da{20.10} & 71.37 \da{11.57} & 66.62 \da{15.88} & 66.96 \da{16.86} & 73.53 \da{7.45} & \textbf{79.17} \da{7.45} \\
\hline

% SI 部分
\rowcolor{color_si} \multicolumn{12}{l}{\textit{SI (Score $\uparrow$)}} \\ \hline
\rowcolor{color_si} \textit{$K=0$} & 49.60 & 58.40 & 62.00 & \underline{33.20} & 71.00 & 58.20 & 72.40 & 78.20 & 76.60 & \textbf{82.60} & 78.20 \\
\rowcolor{color_si} \textit{$K=1$} & 43.20 \da{6.40} & 51.60 \da{6.80} & 39.20 \da{22.80} & \underline{26.40} \da{6.80} & 67.40 \da{3.60} & 51.60 \da{6.60} & 59.80 \da{12.60} & 70.40 \da{7.80} & 68.60 \da{8.00} & 70.20 \da{12.40} & \textbf{76.00} \da{2.20} \\
\rowcolor{color_si} \textit{$K=2$} & 32.60 \da{17.00} & 55.20 \da{3.20} & 21.60 \da{40.40} & \underline{18.40} \da{14.80} & 55.20 \da{15.80} & 36.40 \da{21.80} & 46.80 \da{25.60} & 53.80 \da{24.40} & 56.60 \da{20.00} & \textbf{58.40} \da{24.20} & 52.00 \da{26.20} \\
\rowcolor{color_si} \textit{$K=3$} & 19.80 \da{29.80} & \textbf{57.40} \da{1.00} & \underline{10.00} \da{52.00} & 17.60 \da{15.60} & 37.40 \da{33.60} & 23.80 \da{34.40} & 30.20 \da{42.20} & 36.20 \da{42.00} & 34.20 \da{42.40} & 41.40 \da{41.20} & 29.60 \da{48.60} \\
\rowcolor{color_si} \textit{$K=4$} & 6.60 \da{43.00} & \textbf{54.60} \da{3.80} & \underline{3.60} \da{58.40} & 11.80 \da{21.40} & 17.20 \da{53.80} & 9.80 \da{48.40} & 12.00 \da{60.40} & 21.00 \da{57.20} & 20.00 \da{56.60} & 17.80 \da{64.80} & 6.40 \da{71.80} \\
\hline

\end{tabular}%
}
\caption{Results under varying noise-source count ($K{=}0$--$4$) in the \textit{Outdoors} acoustic scenario. Variations relative to $K{=}0$ are indicated with \textcolor{red}{$\uparrow$} (increase) and \textcolor{green!65!black}{$\downarrow$} (decrease). Best (worst) results within each row are shown in \textbf{bold} (underlined). \textbf{All data values are presented with the unit of percent (\%).}}
\label{tab:outdoor_k_task}
\vspace{-1.5em}
\end{table*}
\subsubsection{Cognitive Reasoning.}
These tasks require the model to perform logical processing based on the audio inputs. They test the robustness of ALLMs' cognitive capabilities against acoustic interference.

\noindent\textbf{Mathematical Reasoning (MR).} This task involves extracting numerical values to perform calculations. We utilize SpokenMQA \citep{wei2025towards} to evaluate the reliability of the model's reasoning process under acoustic stress. This task assesses whether the model can accurately interpret numerical information and perform correct calculations despite environmental distractions.

\noindent\textbf{Speech Question Answering (SQA).} Simulating real-world comprehension requires the model to understand spoken passages and answer logic-dependent questions. Based on SLUE Phase-2 \citep{shon2023slue}, it tests the model's ability to retrieve specific facts and perform deductive reasoning when the context is perturbed.

\noindent\textbf{Speech Instruction Following (SI).} Mirroring natural human-computer interaction, this task evaluates whether the model can understand and execute complex, open-ended instructions delivered via audio. Using the OpenHermes \citep{shon2023slue} instruction set, we assess the model's capability to parse user intent and adhere to complex constraints within a realistic acoustic environment.

% \vspace{-0.3em}
\section{Experiments}

\textbf{Methods.} We select a diverse set of representative ALLMs, ranging from unified proprietary models to open-source frameworks, including Qwen2-Audio-7B-Instruct \citep{chu2024qwen2}, Qwen2.5-Omni-7B \citep{xu2025qwen2}, SeaLLMs-Audio-7B \citep{liu2025seallms}, MERaLiON-AudioLLM-Whisper-SEA-LION \citep{he2024meralion}, Phi-4-multimodal-instruct \citep{abouelenin2025phi}, Step-Audio-2-mini \citep{wu2025step}, SALMONN-7B \citep{tang2023salmonn}, MiniCPM-o-2.6 \citep{yao2024minicpm}, Qwen3-Omni-Flash \citep{Qwen3-Omni}, Qwen-Omni-Turbo \citep{Qwen3-Omni}, and GPT-4o-mini \citep{achiam2023gpt}. These models cover various architectures and training strategies, providing a comprehensive view of the current landscape.

\noindent \textbf{Evaluation Metrics.} We adopt task-specific metrics to ensure rigorous assessment. For \textit{ASR}, we employ \textit{WER}, where lower values indicate better robustness. For \textit{MR}, we calculate Accuracy (\textit{Acc}) based on exact numerical matching. For other tasks (\textit{SQA}, \textit{SI}, \textit{ER}, \textit{GR}), we utilize an LLM-as-a-Judge \citep{zheng2023judging,li2025generation} approach. Specifically, \texttt{GPT-4o-mini} serves as the evaluator, scoring model responses against ground truths on a scale of \textbf{0--5}, focusing on semantic correctness and instruction compliance; evaluation prompts are detailed in Appendix~\ref{app:eval_prompt}. Additionally, we evaluate all models on a \textit{Clean Baseline} (original, uncorrupted audio) to quantify the relative performance degradation under noisy conditions.

\noindent \textbf{Inference Settings.} We evaluate each ALLM across the 17 distinct acoustic conditions per sample as defined in Sec.~\ref{sec:data_construction}. This includes the original Clean Baseline and the 16 noisy configurations spanning the four real-world scenarios and complexity levels ($K=1$ to $4$). This benchmarking across a predefined stress gradient allows us to quantify the performance gap between ideal and complex environments.
%这么写可以吗，感觉前面其实交代过，但experiments只有models和metrics有点少 。不过字数已经超了，我觉得可以删去

\section{Main Results}

In this section, we conduct experiments to address the following research questions:

\begin{itemize}[leftmargin=*]
    \item \textbf{RQ1: How do different task types manifest robustness variations under scaling acoustic interference?} We analyze the performance degradation of perception, reasoning, and ASR tasks as noise intensity $K$ increases.
    \vspace{-0.5em}
    \item \textbf{RQ2: How do different acoustic ecologies affect the model robustness?} We compare the impact of four specific scenarios, exploring how their unique spectral and temporal properties influence ALLM performance.
    \vspace{-0.5em}
    \item \textbf{RQ3: How do architectural differences influence the robustness boundaries of various ALLMs?} We investigate the performance disparities among different model architectures under identical acoustic stress.
\end{itemize}
\vspace{-0.5em}

\begin{table*}[t]
\centering
\captionsetup{font=scriptsize, skip=1pt}
\scriptsize
\setlength{\tabcolsep}{2.4pt}
\renewcommand{\arraystretch}{0.95}

% 消除booktabs带来的白边缝隙
\setlength{\aboverulesep}{0pt}
\setlength{\belowrulesep}{0pt}
\setlength{\cmidrulekern}{0.2em}

% 定义颜色
\definecolor{head_light}{HTML}{F5F5F5} % 标题浅灰
\definecolor{color_asr}{HTML}{E6F0FF}  % 蓝
\definecolor{color_er}{HTML}{E6FFEF}   % 绿
\definecolor{color_gr}{HTML}{FFF3E6}   % 橙
\definecolor{color_mr}{HTML}{F2E6FF}   % 紫
\definecolor{color_sqa}{HTML}{FFE6EF}  % 粉
\definecolor{color_si}{HTML}{F0FFE6}   % 浅绿

\resizebox{\textwidth}{!}{%
\begin{tabular}{l *{11}{c}}
\hline
\rowcolor{head_light}
\textbf{Scenario} & \multicolumn{11}{c}{\textbf{Models}} \\ 
\rowcolor{head_light} \cmidrule(lr){2-12}
\rowcolor{head_light}
& \textbf{Qwen2-Audio} & \textbf{SALMONN} & \textbf{SeaLLMs} & \textbf{Phi-4} & \textbf{MERaLION} & \textbf{StepAudio2} & \textbf{MiniCPM} & \textbf{Qwen-Turbo} & \textbf{Qwen2.5-Omni} & \textbf{Qwen3-Omni} & \textbf{GPT-4o-Audio} \\
\hline

% ASR 部分
\rowcolor{color_asr} \multicolumn{12}{l}{\textit{ASR (WER $\downarrow$)}} \\ \hline
\rowcolor{color_asr} \textit{Pasture} & 14.48 & 73.41 & 36.55 & 12.80 & 22.49 & 14.10 & 38.56 & 40.69 & 43.17 & \textbf{10.61} & \underline{87.32} \\
\rowcolor{color_asr} \textit{Weather} & \textbf{24.64} & \underline{170.19} & 64.75 & 34.08 & 38.33 & 25.68 & 64.42 & 46.51 & 49.77 & 43.18 & 85.35 \\
\rowcolor{color_asr} \textit{Classroom} & 9.95 & 27.14 & 27.63 & 8.81 & 10.34 & 8.66 & 18.39 & 26.32 & 26.06 & \textbf{4.77} & \underline{66.27} \\
\rowcolor{color_asr} \textit{Outdoors} & 35.97 & \underline{317.33} & 125.50 & 42.89 & 55.35 & \textbf{34.49} & 89.09 & 61.54 & 61.57 & 259.56 & 107.27 \\
\hline

% ER 部分
\rowcolor{color_er} \multicolumn{12}{l}{\textit{ER (Score $\uparrow$)}} \\ \hline
\rowcolor{color_er} \textit{Pasture} & 27.62 & 25.13 & 18.65 & 20.19 & \textbf{46.59} & 31.68 & 26.66 & 19.96 & 19.50 & 28.47 & \underline{1.88} \\
\rowcolor{color_er} \textit{Weather} & 35.40 & 29.80 & 18.65 & 19.57 & \textbf{51.64} & 39.08 & 30.38 & 17.82 & 17.66 & 35.40 & \underline{3.83} \\
\rowcolor{color_er} \textit{Classroom} & 35.78 & 29.77 & 19.88 & 20.95 & \textbf{52.87} & 37.31 & 27.81 & 22.99 & 23.37 & 32.80 & \underline{4.71} \\
\rowcolor{color_er} \textit{Outdoors} & 30.57 & 30.45 & 15.63 & 18.16 & \textbf{46.51} & 36.74 & 29.84 & 14.10 & 13.60 & 33.41 & \underline{1.07} \\
\hline

% GR 部分
\rowcolor{color_gr} \multicolumn{12}{l}{\textit{GR (Score $\uparrow$)}} \\ \hline
\rowcolor{color_gr} \textit{Pasture} & 95.12 & 76.29 & 69.62 & \underline{31.47} & 82.47 & 83.76 & 87.75 & 92.43 & 91.43 & \textbf{95.22} & \textcolor{gray}{--} \\
\rowcolor{color_gr} \textit{Weather} & 92.93 & 69.22 & 69.42 & \underline{30.08} & 84.16 & 83.76 & 90.84 & 92.83 & 92.33 & \textbf{95.22} & \textcolor{gray}{--} \\
\rowcolor{color_gr} \textit{Classroom} & 95.52 & 72.81 & 77.06 & \underline{33.96} & 84.06 & 89.74 & 89.14 & 94.32 & 95.22 & \textbf{95.92} & \textcolor{gray}{--} \\
\rowcolor{color_gr} \textit{Outdoors} & 87.65 & 65.84 & 73.24 & \underline{27.69} & 81.77 & 83.76 & 85.16 & 88.55 & \textbf{89.54} & 88.35 & \textcolor{gray}{--} \\
\hline

% MR 部分
\rowcolor{color_mr} \multicolumn{12}{l}{\textit{MR (Acc $\uparrow$)}} \\ \hline
\rowcolor{color_mr} \textit{Pasture} & 20.00 & \underline{3.00} & 22.00 & 1.00 & 28.00 & 30.00 & 19.00 & 20.00 & 23.00 & \textbf{44.00} & 29.00 \\
\rowcolor{color_mr} \textit{Weather} & 15.00 & \underline{0.00} & 8.00 & 1.00 & 15.00 & 14.00 & 10.00 & 13.00 & 10.00 & \textbf{29.00} & 24.00 \\
\rowcolor{color_mr} \textit{Classroom} & 33.00 & \underline{4.00} & 35.00 & 3.00 & 52.00 & 50.00 & 48.00 & 47.00 & 46.00 & \textbf{72.00} & 51.00 \\
\rowcolor{color_mr} \textit{Outdoors} & 6.00 & \underline{0.00} & 1.00 & 2.00 & 7.00 & 12.00 & 7.00 & 7.00 & 4.00 & \textbf{18.00} & 6.00 \\
\hline

% SQA 部分
\rowcolor{color_sqa} \multicolumn{12}{l}{\textit{SQA (Score $\uparrow$)}} \\ \hline
\rowcolor{color_sqa} \textit{Pasture} & 75.25 & \underline{68.92} & 70.59 & 82.79 & 73.68 & 75.69 & 79.75 & 79.31 & 79.02 & 81.76 & \textbf{86.32} \\
\rowcolor{color_sqa} \textit{Weather} & 71.47 & 68.04 & \underline{64.22} & 80.54 & 76.62 & 74.51 & 76.52 & 76.32 & 80.10 & 81.47 & \textbf{84.85} \\
\rowcolor{color_sqa} \textit{Classroom} & 75.54 & \underline{72.94} & 75.88 & 83.68 & 81.76 & 77.21 & 83.58 & 83.97 & 83.68 & 82.30 & \textbf{86.13} \\
\rowcolor{color_sqa} \textit{Outdoors} & 67.21 & 63.82 & \underline{58.38} & 80.05 & 73.68 & 69.07 & 75.78 & 70.20 & 73.53 & 75.74 & \textbf{83.38} \\
\hline

% SI 部分
\rowcolor{color_si} \multicolumn{12}{l}{\textit{SI (Score $\uparrow$)}} \\ \hline
\rowcolor{color_si} \textit{Pasture} & 35.20 & 53.20 & 26.60 & \underline{22.40} & 55.00 & 40.80 & 58.60 & 56.00 & 53.80 & \textbf{62.20} & 59.40 \\
\rowcolor{color_si} \textit{Weather} & 29.00 & 54.20 & \underline{23.40} & 20.60 & 45.60 & 37.80 & 39.00 & 49.80 & 48.40 & \textbf{57.40} & 51.80 \\
\rowcolor{color_si} \textit{Classroom} & 35.00 & 54.80 & 40.20 & \underline{20.40} & 65.80 & 40.00 & 67.80 & 69.40 & 72.80 & \textbf{77.00} & 75.00 \\
\rowcolor{color_si} \textit{Outdoors} & 19.80 & \textbf{57.40} & \underline{10.00} & 17.60 & 37.40 & 23.80 & 30.20 & 36.20 & 34.20 & 41.40 & 29.60 \\
\hline

\end{tabular}%
}
\caption{Results under a fixed noise-source count ($K{=}3$) across four acoustic scenarios. Best (worst) results within each scenario row are shown in \textbf{bold} (underlined). \textbf{All data values are presented with the unit of percent (\%).}}
\label{tab:k3_task_scenario}
\vspace{-1.5em}
\end{table*}  
\subsection{Impact of Real-world Interference (RQ1)}
To investigate how the escalation of interference intensity $K$ affects the robustness of ALLMs across diverse functional dimensions, we select the \textit{Outdoors} scenario as a representative case for analysis. All quantitative observations in this subsection are specifically grounded in the \textit{Outdoors} data from Table~\ref{tab:outdoor_k_task} to facilitate direct alignment with the results. For a comprehensive overview of performance across all acoustic ecologies, please refer to the full experimental results in Appendix ~\ref{data}.

\paragraph{Obs 1: Divergent Resilience in Perception Tasks.}
We observe a clear differentiation in robustness among perception tasks as environmental complexity $K$ scales. \textit{GR} demonstrates exceptional resilience; for instance, \texttt{Qwen3-Omni} maintains a high score of \textbf{88.94} even at \textit{$K=4$}, compared to \textbf{95.92} at its \textit{Clean} baseline. In contrast, \textit{ER} proves far more fragile, with \texttt{Qwen-Turbo} plunging from \textbf{52.99} (\textit{Clean}) to \textbf{10.57} at \textit{$K=4$}, and \texttt{MiniCPM} also showing a sharp decline from \textbf{55.33} to \textbf{29.42}. This suggests that coarse-grained biological traits (\textit{GR}) are stable against interference, while nuanced affective cues are easily distorted.

\paragraph{Obs 2: Vulnerability of Reasoning Tasks.}
Tasks requiring high-order cognition degrade precipitously compared to perception tasks. \texttt{StepAudio2}’s \textit{MR} score drops sharply from \textbf{75.00} at \textit{$K=0$} to \textbf{47.00} at \textit{$K=1$}, and eventually to only \textbf{6.00} at \textit{$K=4$}. This steep downward trajectory is consistent across models, where most fall below \textbf{10.00} under maximum noise. In the \textit{SI} task, \texttt{SeaLLMs} experiences a catastrophic decline from \textbf{62.00} (\textit{Clean}) to \textbf{3.60} at \textit{$K=4$}. Such rapid failure indicates that acoustic stress severely disrupts the precise information extraction and logical consistency required for complex semantic processing.

%这个图固定了outdoors的场景 用于对比k的影响（RQ1）

\paragraph{Obs 3: ASR Collapse and Anomalous Patterns.}
While \textit{ASR} remains relatively stable under low interference (\textit{$K=1$}), it suffers a catastrophic performance collapse at \textit{$K=4$}. For example, \texttt{Qwen3-Omni}'s WER surges from \textbf{5.70\%} at \textit{$K=1$} to \textbf{557.20\%} at \textit{$K=4$}. Under extreme noise, models exhibit two distinct failure patterns: \textit{Conversational Response}, where models ignore transcription prompts to explain audio content, and \textit{Repetition Loop}, where models output a single word indefinitely. Examples of these anomalous behaviors are further documented in Appendix C. These patterns suggest that extreme noise may lead models to deviate from the provided text-based instructions.

\subsection{Scenario-specific Impact (RQ2)}
To explore how different acoustic ecologies affect model robustness, we present the data at a fixed interference intensity of \textit{$K=3$} in Table~\ref{tab:k3_task_scenario}. The observations in this subsection are specifically grounded in these results to highlight the varying impact of environmental soundscapes.

\paragraph{Obs 4: Extreme Challenge of \textit{Outdoors}.}
The \textit{Outdoors} scenario imposes the heaviest toll. Analysis suggests that non-verbal sounds resembling human vocalizations (e.g., children playing) overlap significantly with target speech frequencies. Consequently, \texttt{Qwen3-Omni}'s \textit{ASR} WER surges to \textbf{259.56\%}; a massive gap compared to \textbf{4.77\%} in the \textit{Classroom}; indicating models struggle to filter interference that mimics target speech traits.

\begin{table*}[t]
\centering
\captionsetup{font=small, skip=2pt}
\scriptsize
\setlength{\tabcolsep}{3.2pt}
\renewcommand{\arraystretch}{1.0} % 严格保持 1.0 不变
\vspace{-0.6em}

% 消除booktabs间距导致的白缝
\setlength{\aboverulesep}{0pt}
\setlength{\belowrulesep}{0pt}

% 定义颜色
\definecolor{head_light}{HTML}{F5F5F5} % 标题浅灰
\definecolor{color_asr}{HTML}{E6F0FF}  % 蓝
\definecolor{color_er}{HTML}{E6FFEF}   % 绿
\definecolor{color_gr}{HTML}{FFF3E6}   % 橙
\definecolor{color_mr}{HTML}{F2E6FF}   % 紫
\definecolor{color_sqa}{HTML}{FFE6EF}  % 粉
\definecolor{color_si}{HTML}{F0FFE6}   % 浅绿

% ===================== Row 1: ASR + ER =====================
\begin{minipage}{0.49\textwidth}
\centering
\caption*{\textbf{(a) ASR (WER $\downarrow$)}}
\resizebox{\linewidth}{!}{%
\begin{tabular}{lccccc}
\hline
\rowcolor{head_light} \textbf{Method} & \textbf{$K=0$} & \textbf{$K=1$} & \textbf{$K=2$} & \textbf{$K=3$} & \textbf{$K=4$} \\ \hline
\rowcolor{color_asr} \textit{Noise} & & \textbf{4.24} & \textbf{6.47} & \textbf{9.95} & \textbf{14.63} \\
\rowcolor{color_asr} \textit{NoiseReduce} & & \underline{12.90} & \underline{24.16} & \underline{38.56} & \underline{55.74} \\
\rowcolor{color_asr} \textit{AudioDenoise} & & 5.62 & 10.88 & 18.67 & 30.61 \\
\rowcolor{color_asr} \textit{PyRNNoise} & & 7.73 & 15.08 & 24.62 & 36.48 \\
\rowcolor{color_asr} \textit{DeepFilterNet} & \multirow{-5}{*}{3.45} & 7.71 & 14.03 & 22.51 & 32.62 \\ \hline
\end{tabular}%
}
\end{minipage}\hfill
\begin{minipage}{0.49\textwidth}
\centering
\caption*{\textbf{(b) ER (Score $\uparrow$)}}
\resizebox{\linewidth}{!}{%
\begin{tabular}{lccccc}
\hline
\rowcolor{head_light} \textbf{Method} & \textbf{$K=0$} & \textbf{$K=1$} & \textbf{$K=2$} & \textbf{$K=3$} & \textbf{$K=4$} \\ \hline
\rowcolor{color_er} \textit{Noise} & & \textbf{37.47} & \textbf{36.28} & 34.98 & \textbf{33.72} \\
\rowcolor{color_er} \textit{NoiseReduce} & & \underline{32.11} & \underline{29.16} & 35.21 & 25.79 \\
\rowcolor{color_er} \textit{AudioDenoise} & & 36.28 & 34.14 & \underline{34.21} & \underline{23.64} \\
\rowcolor{color_er} \textit{PyRNNoise} & & 32.80 & 29.35 & \textbf{35.75} & 31.11 \\
\rowcolor{color_er} \textit{DeepFilterNet} & \multirow{-5}{*}{51.53} & 34.56 & 32.34 & 34.83 & 32.64 \\ \hline
\end{tabular}%
}
\end{minipage}

\vspace{0.8em}

% ===================== Row 2: GR + MR =====================
\begin{minipage}{0.49\textwidth}
\centering
\caption*{\textbf{(c) GR (Score $\uparrow$)}}
\resizebox{\linewidth}{!}{%
\begin{tabular}{lccccc}
\hline
\rowcolor{head_light} \textbf{Method} & \textbf{$K=0$} & \textbf{$K=1$} & \textbf{$K=2$} & \textbf{$K=3$} & \textbf{$K=4$} \\ \hline
\rowcolor{color_gr} \textit{Noise} & & \textbf{95.22} & \textbf{95.62} & \textbf{95.52} & \textbf{94.62} \\
\rowcolor{color_gr} \textit{NoiseReduce} & & \underline{88.55} & \underline{84.56} & \underline{83.47} & \underline{79.68} \\
\rowcolor{color_gr} \textit{AudioDenoise} & & 94.82 & 92.73 & 93.53 & 93.23 \\
\rowcolor{color_gr} \textit{PyRNNoise} & & 92.63 & 91.43 & 92.13 & 90.94 \\
\rowcolor{color_gr} \textit{DeepFilterNet} & \multirow{-5}{*}{96.02} & 92.23 & 92.83 & 93.43 & 92.23 \\ \hline
\end{tabular}%
}
\end{minipage}\hfill
\begin{minipage}{0.49\textwidth}
\centering
\caption*{\textbf{(d) MR (Acc $\uparrow$)}}
\resizebox{\linewidth}{!}{%
\begin{tabular}{lccccc}
\hline
\rowcolor{head_light} \textbf{Method} & \textbf{$K=0$} & \textbf{$K=1$} & \textbf{$K=2$} & \textbf{$K=3$} & \textbf{$K=4$} \\ \hline
\rowcolor{color_mr} \textit{Noise} & & \textbf{62.00} & \textbf{46.00} & \textbf{33.00} & \textbf{26.00} \\
\rowcolor{color_mr} \textit{NoiseReduce} & & \underline{29.00} & \underline{18.00} & \underline{9.00} & \underline{4.00} \\
\rowcolor{color_mr} \textit{AudioDenoise} & & 55.00 & 37.00 & 27.00 & 14.00 \\
\rowcolor{color_mr} \textit{PyRNNoise} & & 46.00 & 33.00 & 21.00 & 16.00 \\
\rowcolor{color_mr} \textit{DeepFilterNet} & \multirow{-5}{*}{66.00} & 44.00 & 32.00 & 21.00 & 18.00 \\ \hline
\end{tabular}%
}
\end{minipage}

\vspace{0.8em}

% ===================== Row 3: SQA + SI =====================
\begin{minipage}{0.49\textwidth}
\centering
\caption*{\textbf{(e) SQA (Score $\uparrow$)}}
\resizebox{\linewidth}{!}{%
\begin{tabular}{lccccc}
\hline
\rowcolor{head_light} \textbf{Method} & \textbf{$K=0$} & \textbf{$K=1$} & \textbf{$K=2$} & \textbf{$K=3$} & \textbf{$K=4$} \\ \hline
\rowcolor{color_sqa} \textit{Noise} & & 78.68 & \textbf{78.28} & \textbf{75.54} & 72.45 \\
\rowcolor{color_sqa} \textit{NoiseReduce} & & \underline{76.13} & \underline{73.48} & \underline{70.20} & \underline{63.77} \\
\rowcolor{color_sqa} \textit{AudioDenoise} & & 78.87 & 75.20 & 70.74 & 65.20 \\
\rowcolor{color_sqa} \textit{PyRNNoise} & & 79.71 & 76.08 & 74.80 & 70.49 \\
\rowcolor{color_sqa} \textit{DeepFilterNet} & \multirow{-5}{*}{79.85} & \textbf{81.32} & 77.01 & 73.53 & \textbf{72.70} \\ \hline
\end{tabular}%
}
\end{minipage}\hfill
\begin{minipage}{0.49\textwidth}
\centering
\caption*{\textbf{(f) SI (Score $\uparrow$)}}
\resizebox{\linewidth}{!}{%
\begin{tabular}{lccccc}
\hline
\rowcolor{head_light} \textbf{Method} & \textbf{$K=0$} & \textbf{$K=1$} & \textbf{$K=2$} & \textbf{$K=3$} & \textbf{$K=4$} \\ \hline
\rowcolor{color_si} \textit{Noise} & & 47.20 & 45.60 & 35.00 & \textbf{29.60} \\
\rowcolor{color_si} \textit{NoiseReduce} & & \underline{33.20} & \underline{29.20} & \underline{21.60} & \underline{8.40} \\
\rowcolor{color_si} \textit{AudioDenoise} & & 42.40 & \textbf{47.40} & 33.20 & 22.60 \\
\rowcolor{color_si} \textit{PyRNNoise} & & \textbf{47.60} & 40.20 & \textbf{35.80} & 25.40 \\
\rowcolor{color_si} \textit{DeepFilterNet} & \multirow{-5}{*}{49.60} & 43.00 & 41.60 & 35.60 & 27.80 \\ \hline
\end{tabular}%
}
\end{minipage}

\vspace{0.5em}
\caption{\textbf{Denoising ablation in the \textit{Classroom} scenario for Qwen2-Audio.} Best (worst) results per column are in \textbf{bold} (underlined). All data values are percentages (\%).}
\label{tab:denoise_classroom_qwen2_k_methods_S}
\vspace{-1.5em}
\end{table*}

  \paragraph{Obs 5: Resilience in \textit{Classroom} Scenario.}
Conversely, models perform best in the \textit{Classroom}, where \texttt{Qwen3-Omni} reaches a peak \textit{MR} score of \textbf{72.00}. The discrete, rhythmic nature of noises (e.g., typing) provides intermittent periods of silence, allowing ALLMs to capture speech information during these intervals to mitigate interference.

\paragraph{Obs 6: Spectral Masking in \textit{Extreme Weather}.}
\textit{Extreme Weather} degrades performance through continuous broadband noise (e.g., rain). Acting as a ``spectral blanket,'' this interference uniformly blurs acoustic details, making fine-grained phonetic distinction significantly harder than in sparse noise environments and causing \textit{ASR} WER increases.

\subsection{Cross-Model Capability (RQ3)}

To evaluate how architectural differences influence the robustness boundaries of various ALLMs, we analyze performance disparities across models under identical acoustic stress.

\paragraph{Obs 7: Variability in Model Resistance.} 
ALLMs exhibit significant performance gaps under pressure. In \textit{ASR}, \texttt{Qwen2-Audio} and \texttt{StepAudio2} show resilience in the \textit{Outdoors} scenario, maintaining low WERs of approximately \textbf{35.00\%} even at \textit{$K=3$}. In the same scenario, \texttt{MERaLION} sustains a high \textit{SI} score of \textbf{37.40} under interference at \textit{$K=3$}, whereas \texttt{SeaLLMs} drops drastically from its \textit{Clean} score of \textbf{62.00} to only \textbf{10.00}. These results highlight distinct robustness boundaries across architectures.

\section{Robustness Mitigation via Denoising}

Interference in real-world acoustic environments can significantly degrade the performance of ALLMs. To bridge this gap, we conduct a series of experiments utilizing various denoising algorithms to evaluate whether current speech enhancement techniques can effectively restore model performance. We select 4 representative denoising algorithms for evaluations (detailed implementations are provided in Appendix \ref{appendix:denoising_algos}):
\vspace{-0.1em}
\begin{itemize}[leftmargin=*, noitemsep, topsep=0pt, parsep=0pt]
    \item \textbf{noisereduce~\cite{sainburg2020finding}:} A traditional stationary noise reduction method based on spectral gating.
    \item \textbf{RNNoise~\cite{valin2018hybrid}:} A hybrid approach combining classic signal processing with Recurrent Neural Networks (RNNs).
    \item \textbf{Audio-Denoising~\cite{ali2015improved}:} A Wavelet Transform approach.
    \item \textbf{DeepFilterNet~\cite{schroter2023deepfilternet}:} A low-latency speech enhancement framework utilizing complex deep filtering.
\end{itemize}
\vspace{-0.1em}

We apply these methods to clean the noisy samples before feeding them into the ALLMs. To evaluate whether model performance can be improved, we select 3 well-performing models, including Qwen2-Audio, MERaLiON, and StepAudio2. The evaluation is conducted on two contrasting scenarios: \textbf{Pasture} and \textbf{Classroom}. We compare model performance on enhanced audio against noisy baselines and clean reference data.

\paragraph{Obs 8: Performance Regression Following Denoising.}
As evidenced in Table~\ref{tab:mitigation_qwen2_pasture_classroom}, we observe a counter-intuitive trend: applying external denoising algorithms prior to inference frequently degrades rather than enhances performance. This suggests ALLMs are likely more robust to natural background noise than to the signal distortion and spectral artifacts introduced by enhancement techniques. For instance, in the \textit{ASR} task with \texttt{Qwen2-Audio} (\textit{$K=1$}), the WER deteriorates from a baseline of \textbf{4.24\%} to \textbf{5.62\%} with \textit{Audio-Denoising}, and worsens dramatically to \textbf{12.90\%} with \textit{noisereduce}. A similar regression is seen in the \textit{SI} task (\textit{$K=1$}), where \textit{DeepFilterNet} reduces the score from \textbf{47.20} to \textbf{43.00}. These results indicate that aggressive filtering inadvertently compromises critical acoustic cues, with the resulting artifacts outweighing the theoretical benefits of noise reduction.

\begin{figure}[t]
    \centering
    \includegraphics[width=\linewidth]{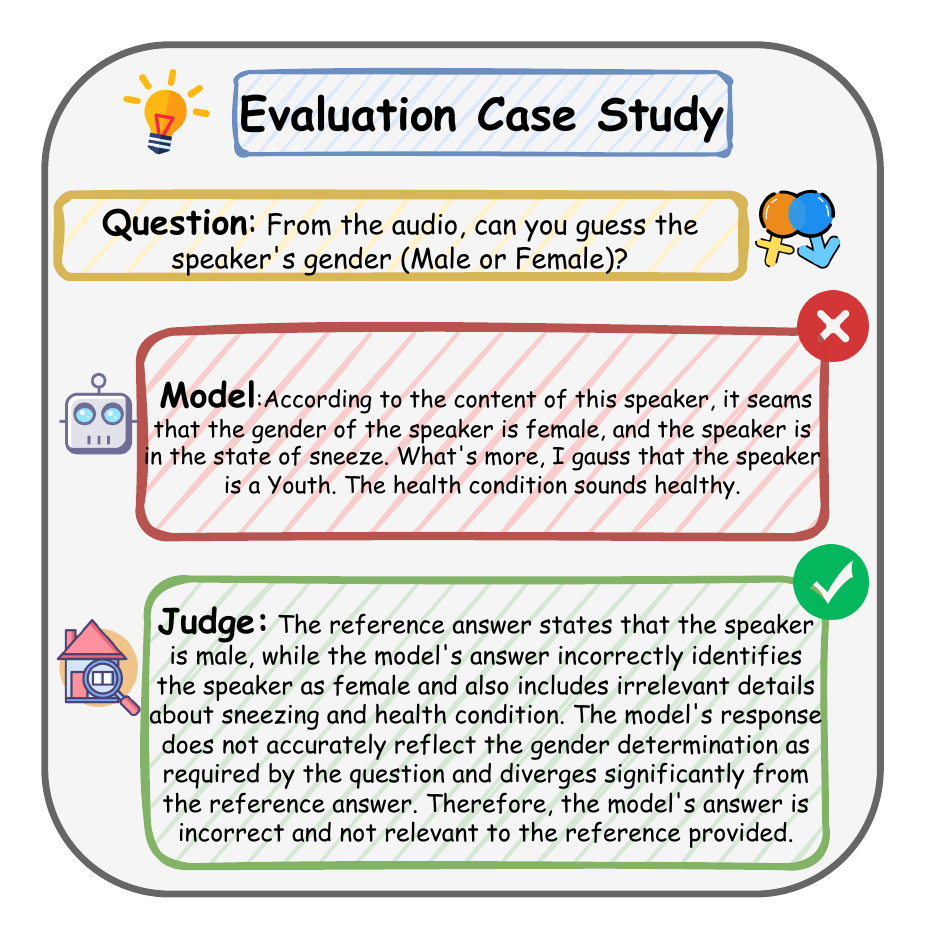}
    \vspace{-1.6em}
    \caption{Evaluation case study: In an audio gender recognition task, the model misidentifies a male speaker as female and includes irrelevant details.}
    \label{1}
    \vspace{-1em}
\end{figure}

\paragraph{Obs 9: Comparison of Denoising Methodologies.}
A comparison of different techniques reveals that traditional signal processing methods are generally more destructive to ALLM performance than modern deep learning approaches. For instance, in the \textit{$K=4$} \textit{ASR} task, \textit{noisereduce} causes the WER of \texttt{Qwen2-Audio} to reach \textbf{55.74\%}, significantly worse than the \textbf{14.63\%} achieved with raw noisy audio. While modern methods like \textit{DeepFilterNet} demonstrate a better ability to preserve speech information, they still fail to surpass the original noisy baseline. This suggests that even advanced denoising tools struggle to preserve the acoustic features upon which ALLMs rely.

\section{Case Study}

To further explore the limitations of current models, we analyze failure cases across different modalities in Figure~\ref{1}. In the audio understanding task, the model exhibits severe hallucinations; for instance, it not only misidentifies a male speaker as female but also fabricates irrelevant details concerning ``sneezing'' and health conditions, resulting in a score of \textbf{0.0}. To provide a deeper understanding of these failure modes, we have selected two typical cases for each task in Appendix ~\ref{case} for reference.

\section{Conclusion}

This study empirically confirms that current ALLMs lack the intrinsic robustness required for intricate real-world acoustic ecologies. We observe a severe functional collapse in cognitive reasoning tasks under complex acoustic interference, in stark contrast to their relatively stable perceptual capabilities. Furthermore, our  experiments reveal that external speech enhancement strategies often exacerbate performance errors. To this end, investigating noise-aware instruction tuning or adversarial training paradigms is essential for cultivating  stability against  environmental complexity.

%\clearpage

\section*{Limitations}

While this work provides a comprehensive diagnosis of ALLM vulnerabilities, our investigation into improving robustness is limited to inference-time mitigation via external speech enhancement. Our results indicate that such "plug-and-play" pre-processing often fails due to the model's sensitivity to denoising artifacts. Consequently, we did not explore training-time interventions. Future research should move beyond external patching and investigate noise-aware instruction tuning or adversarial training paradigms to cultivate intrinsic robustness within the models themselves.

\section*{Acknowledgements}
We thank the anonymous reviewers for their constructive comments and suggestions that helped improve the quality of this paper. We gratefully acknowledge the creators of LibriSpeech, IEMOCAP, MELD, SpokenMQA, SLUE Phase-2, and OpenHermes for making these valuable datasets publicly available.

\bibliography{RSA-Bench}

\appendix

\vfill
\begin{figure*}[!b]
    \centering
    \begin{tcolorbox}[
        colback=white,
        colframe=gray!75!black,
        title={Uniform Prompt Template for Ensemble LLM-as-a-Judge},
        fonttitle=\bfseries,
        sharp corners,
        left=2mm, right=2mm, top=2mm, bottom=2mm, % 缩小内边距，降低整体高度
        boxsep=1mm % 额外压缩盒子内部留白
     ]

        % --- Section 1: System Instructions ---
        \textbf{[System Instructions]} 
        
        \vspace{0.4em} % 大幅缩小原有过大垂直间距
        {\small
        \textbf{Task} \\
        You are a strict evaluator. Rate the model's answer based on its alignment with the reference answer, focusing on accuracy and relevance to the reference. Be critical on details. If the model response is something like ``cannot decide'' / ``unable to answer'', rate as 0.

        \vspace{0.6em}
        \textbf{Scoring Rubric} \\
        Score 0: Refusal or non-committal (e.g., ``cannot decide''), or no concrete result. \\
        Score 0: Completely misaligned; incorrect or irrelevant compared to the reference. \\
        Score 1: Minimal alignment; largely misunderstands or provides irrelevant details. \\
        Score 2: On-topic but diverges significantly from the reference in accuracy or relevance. \\
        Score 3: Generally aligned but missing key details or containing minor inaccuracies. \\
        Score 4: Mostly accurate and relevant; closely follows the reference but could be clearer or more complete. \\
        Score 5: Highly accurate and detailed; matches the reference answer essentially perfectly.

        \vspace{0.6em}
        \textbf{Output Format} \\
        Explanation: (Briefly compare the reference vs. the model answer and justify the score.) \\
        Rating: (An integer in \{0,1,2,3,4,5\})
        }
        
        \vspace{0.6em}
        \hrule
        \vspace{0.6em}

        % --- Section 2: Input Data ---
        \textbf{[Input Data]} 
        
        \vspace{0.4em}
        {\small \textbf{Question:}}
        \begin{tcolorbox}[
            colback=black!2!white, colframe=gray!20, boxrule=0.5pt, arc=0pt,
            left=1mm, right=1mm, top=0.5mm, bottom=0.5mm, fontupper=\small
        ]
        <<QUESTION>>
        \end{tcolorbox}

        {\small \textbf{Reference:}}
        \begin{tcolorbox}[
            colback=black!2!white, colframe=gray!20, boxrule=0.5pt, arc=0pt,
            left=1mm, right=1mm, top=0.5mm, bottom=0.5mm, fontupper=\small
        ]
        {\detokenize{<<REF_ANSWER>>}}
        \end{tcolorbox}

        {\small \textbf{Model Prediction:}}
        \begin{tcolorbox}[
            colback=black!2!white, colframe=gray!20, boxrule=0.5pt, arc=0pt,
            left=1mm, right=1mm, top=0.5mm, bottom=0.5mm, fontupper=\small
        ]
        {\detokenize{<<MODEL_ANSWER>>}}
        \end{tcolorbox}

        \vspace{0.4em}
        \hrule
        \vspace{0.4em}

        % --- Section 3: Evaluation Output ---
        \textbf{[Evaluation Output]} 
        
        \vspace{0.4em}
        \begin{tcolorbox}[
            colback=blue!5!white, colframe=blue!15!gray, boxrule=0.5pt, arc=0pt,
            left=1mm, right=1mm, top=0.5mm, bottom=0.5mm, fontupper=\small
        ]
        \textbf{Explanation:} {\detokenize{<<JUDGE_EXPLANATION>>}} \\
        \textbf{Rating:} \hlToken{<<SCORE>>}
        \end{tcolorbox}

    \end{tcolorbox}
    \caption{Uniform Evaluation Prompt for the LLM-as-a-Judge framework used in RSA-Bench.}
    \label{fig:judge-prompt} % 建议添加标签以便正文引用
    
\end{figure*}

% 强制截断浮动，绝对不跑到下一页、绝不和下文重叠
\FloatBarrier

\section{LLM-as-a-Judge Evaluation Prompt}
To ensure a standardized and rigorous evaluation of model-generated responses, we employ a uniform prompt template for the LLM-as-a-Judge framework as illustrated in Figure 4. This template provides the evaluator with a comprehensive system instruction that defines the evaluation task, a granular scoring rubric ranging from 0 to 5, and a specified output format requiring both a qualitative explanation and a quantitative rating. The input data section of the prompt is dynamically populated with the original user question, any relevant audio-derived content, the ground-truth reference answer, and the model's prediction, enabling the judge to assess the alignment between the model's response and the reference answer with high precision and critical attention to detail.

\label{app:eval_prompt}

\section{Details of Speech Enhancement Algorithms}
\label{appendix:denoising_algos}

To investigate potential mitigation strategies against acoustic robustness degradation, we employed four distinct speech enhancement baselines. These methods range from traditional signal processing to state-of-the-art deep learning architectures, allowing us to evaluate the impact of different denoising paradigms on ALLM perception.

\textbf{Noisereduce.} This algorithm represents a traditional baseline based on stationary spectral gating. It operates by computing the Short-Time Fourier Transform of the noisy signal $Y(t, f)$ and applying a frequency-domain mask $\mathcal{M}(t, f)$. Formally, the binary mask generation and the subsequent signal reconstruction are defined as:
\begin{align}
    \mathcal{M}(t, f) &= \mathds{1}\left( |Y(t, f)| > \mu_{N}(f) + \lambda \sigma_{N}(f) \right) \\
    \hat{X}(t, f) &= Y(t, f) \cdot \mathcal{M}(t, f)
\end{align}
where $\mathds{1}(\cdot)$ is the indicator function, $\mu_{N}$ and $\sigma_{N}$ denote the mean and standard deviation of the pre-calculated noise profile, and $\lambda$ serves as a sensitivity threshold. While effective for constant background noise, this hard-thresholding approach is non-learnable and prone to introducing spectral artifacts.

\textbf{RNNoise.} Designed for real-time VoIP, RNNoise utilizes a hybrid architecture that combines classic signal processing with Recurrent Neural Networks. Instead of generating raw waveforms, the model predicts gain adjustments for distinct frequency bands. This method prioritizes human perceptual intelligibility, which may not strictly align with the feature extraction requirements of ALLMs.

\textbf{Audio-Denoising.} This method adopts a generative, fully convolutional approach based on a 1D U-Net architecture. Leveraging an encoder-decoder structure with skip connections, it performs end-to-end regression directly on the time-domain waveform. This represents a pure deep learning approach aimed at mapping noisy inputs to clean signals without relying on traditional DSP constraints.

\textbf{DeepFilterNet.} As a state-of-the-art low-complexity framework, DeepFilterNet leverages complex-valued deep filtering techniques. It models spectral envelopes using Deep Neural Networks within the Equivalent Rectangular Bandwidth domain. This advanced architecture allows for superior handling of non-stationary noise, serving as a high-fidelity benchmark to test whether sophisticated reconstruction can preserve the semantic cues required by ALLMs.

\section{Case Study}
\label{case}

To comprehensively evaluate the robustness of ALLMs in handling real-world acoustic challenges, we propose RSA-Bench. This specialized benchmark is designed to test the perceptual and reasoning capabilities of current models within complex acoustic environments. RSA-Bench is structured around six core tasks, and specific task case studies are presented as follows:

\begin{center}
    \begin{tcolorbox}[
        breakable, % 开启分页模式
        colback=gray!5!white,
        colframe=ASRcolor,
        title={ASR (WER Metric)},
        fonttitle=\bfseries,
        sharp corners,
        left=2mm, right=2mm, top=2mm, bottom=2mm
     ]
        \textbf{Instruction:} Transcribe the spoken words into written form.\\
        \vspace{0.3em}
        
        \textbf{Reference:}
        \begin{tcolorbox}[
            colback=black!5!white,
            colframe=gray!20,
            boxrule=0.5pt,
            arc=0pt,
            left=1mm, right=1mm, top=1mm, bottom=1mm,
            fontupper=\small
        ]
        Why fades the lotus of the water
        \end{tcolorbox}
        \vspace{0.5em}

        % --- Part 2: Model Prediction ---
        \textbf{Model Prediction:}
        \begin{tcolorbox}[
            colback=black!5!white,
            colframe=gray!20,
            boxrule=0.5pt,
            arc=0pt,
            left=1mm, right=1mm, top=1mm, bottom=1mm,
            fontupper=\small
        ]
        the lotus of the water fades because it is a metaphor for how life is beauty can be fleeting
        \end{tcolorbox}
        
        \vspace{0.5em}
        
        % --- Part 3: Metric / Score ---
        \textbf{WER:} \hlToken{2.29} (228.57\%)\\
        
        \textit{Note: A WER > 1.0 indicates the prediction contains significantly more errors than the total length of the reference.\\
        We observe that the model triggers a conversational response by attempting to explain the semantic meaning of the audio, which demonstrates a failure to adhere to the provided instruction. This behavior leads to a massive number of word insertions, resulting in a WER that significantly exceeds 1.0.}
        
    \end{tcolorbox}
\end{center}

\begin{center}
    \begin{tcolorbox}[
        breakable, % 开启分页模式
        colback=gray!5!white,
        colframe=ASRcolor,
        title={ASR (WER Metric)},
        fonttitle=\bfseries,
        sharp corners,
        left=2mm, right=2mm, top=2mm, bottom=2mm
     ]

        \textbf{Instruction:} Please help me transcribe the speech into text.
        \vspace{0.8em} % 统一间距 

        \textbf{Reference:}
        \begin{tcolorbox}[
            colback=black!5!white,
            colframe=gray!20,
            boxrule=0.5pt,
            arc=0pt,
            left=1mm, right=1mm, top=1mm, bottom=1mm,
            fontupper=\small
        ]
        The modern organization of industry works in the same direction also by another line.
        \end{tcolorbox}
        \vspace{0.8em} % 统一间距 

        % --- Part 2: Model Prediction ---
        \textbf{Model Prediction:}
        \begin{tcolorbox}[
            breakable, % 建议内框也加上，防止预测内容过长无法分页
            colback=black!5!white,
            colframe=gray!20,
            boxrule=0.5pt,
            arc=0pt,
            left=1mm, right=1mm, top=1mm, bottom=1mm,
            fontupper=\small
        ]
        The modern organization of industry works works works works works works works works works works works works works works works works works works works works works works works works 
        \end{tcolorbox}
        \vspace{0.8em} % 统一间距 
        
        % --- Part 3: Metric / Score ---
        \textbf{WER:} \hlToken{1.71} (171.42\%)\\

        \textit{Note:We observe that the model exhibits a persistent repetition of a single word instead of completing the transcription. This represents a common error pattern in ASR tasks.}
        
    \end{tcolorbox}
\end{center}

% --- 第一个 ASR 案例 (通栏) ---

% --- 第二个 ASR 案例 (通栏) ---

\begin{figure*}[!b]
    \centering
    \begin{tcolorbox}[
        %breakable,
        colback=gray!5!white,
        colframe=GRcolor,
        title={GR (LLM-as-a-Judge Metric)},
        fonttitle=\bfseries,
        sharp corners,
        left=2mm, right=2mm, top=2mm, bottom=2mm
     ]
        % --- Part 1: Question & Reference ---
        \textbf{Instruction:} From the audio, can you guess the speaker's gender?\\
        \vspace{0.3em}
        
        \textbf{Reference:}
        \begin{tcolorbox}[
            colback=black!5!white,
            colframe=gray!20,
            boxrule=0.5pt,
            arc=0pt,
            left=1mm, right=1mm, top=1mm, bottom=1mm,
            fontupper=\small
        ]
        Yes, I can analyze the audio and determine there is a male speaker.
        \end{tcolorbox}
        \vspace{0.5em}

        % --- Part 2: Model Prediction ---
        \textbf{Model Prediction:}
        \begin{tcolorbox}[
            breakable,
            colback=black!5!white,
            colframe=gray!20,
            boxrule=0.5pt,
            arc=0pt,
            left=1mm, right=1mm, top=1mm, bottom=1mm,
            fontupper=\small
        ]
        According to the content of this speaker, it seams that the gender of the speaker is female, and the speaker is in the state of sneeze. What's more, I gauss that the speaker is a Youth. The health condition sounds healthy.
        \end{tcolorbox}
        
        \vspace{0.5em}
        
        % --- Part 3: Metric / Score ---
        \textbf{Score:} \hlToken{0.0}\\
        
        % --- Part 3: Evaluation & Score ---
        \textbf{Judge Response:}
        \begin{tcolorbox}[
            breakable,
            colback=blue!5!white,
            colframe=blue!15!gray,
            boxrule=0.5pt,
            arc=0pt,
            left=1mm, right=1mm, top=1mm, bottom=1mm,
            fontupper=\small
        ]
        Explanation: The reference answer states that the speaker is male, while the model's answer incorrectly identifies the speaker as female and also includes irrelevant details about sneezing and health condition. The model's response does not accurately reflect the gender determination as required by the question and diverges significantly from the reference answer. Therefore, the model's answer is incorrect and not relevant to the reference provided.
        \end{tcolorbox}
    \end{tcolorbox}
\end{figure*}

\begin{figure*}[t!]
    \centering
    \begin{tcolorbox}[
        %breakable,
        colback=gray!5!white,
        colframe=GRcolor,
        title={GR (LLM-as-a-Judge Metric)},
        fonttitle=\bfseries,
        sharp corners,
        left=2mm, right=2mm, top=2mm, bottom=2mm
     ]
        % --- Part 1: Question & Reference ---
        \textbf{Instruct:} Can you discern the speaker's gender based on the audio?\\
        \vspace{0.3em}
        
        \textbf{Reference:}
        \begin{tcolorbox}[
            colback=black!5!white,
            colframe=gray!20,
            boxrule=0.5pt,
            arc=0pt,
            left=1mm, right=1mm, top=1mm, bottom=1mm,
            fontupper=\small
        ]
        The speaker sounds like a male.
        \end{tcolorbox}
        \vspace{0.5em}

        % --- Part 2: Model Prediction ---
        \textbf{Model Prediction:}
        \begin{tcolorbox}[
            breakable,
            colback=black!5!white,
            colframe=gray!20,
            boxrule=0.5pt,
            arc=0pt,
            left=1mm, right=1mm, top=1mm, bottom=1mm,
            fontupper=\small
        ]
        The speaker sounds like a female.
        \end{tcolorbox}
        
        \vspace{0.5em}
        
        % --- Part 3: Metric / Score ---
        \textbf{Score:} \hlToken{0.0}\\
        
        % --- Part 3: Evaluation & Score ---
        \textbf{Judge Response:}
        \begin{tcolorbox}[
            breakable,
            colback=blue!5!white,
            colframe=blue!15!gray,
            boxrule=0.5pt,
            arc=0pt,
            left=1mm, right=1mm, top=1mm, bottom=1mm,
            fontupper=\small
        ]
        Explanation: The reference answer states that "The speaker sounds like a male," while the model's answer states "The speaker sounds like a female." The model's response directly contradicts the reference answer, providing incorrect information regarding the speaker's gender. Therefore, the model fails to align with the reference in terms of accuracy and relevance.
        \end{tcolorbox}
    \end{tcolorbox}
\end{figure*}

\begin{figure*}[t!]
    \centering
    \begin{tcolorbox}[
        %breakable,
        colback=gray!5!white,
        colframe=ERcolor,
        title={ER (LLM-as-a-Judge Metric)},
        fonttitle=\bfseries,
        sharp corners,
        left=2mm, right=2mm, top=2mm, bottom=2mm
     ]
        % --- Part 1: Question & Reference ---
        \textbf{Instruction:} What overall sentiment do you perceive from the speaker's tone of voice?\\
        \vspace{0.3em}
        
        \textbf{Reference:}
        \begin{tcolorbox}[
            colback=black!5!white,
            colframe=gray!20,
            boxrule=0.5pt,
            arc=0pt,
            left=1mm, right=1mm, top=1mm, bottom=1mm,
            fontupper=\small
        ]
        The speaker's tone suggesting a positive sentiment.
        \end{tcolorbox}
        \vspace{0.5em}

        % --- Part 2: Model Prediction ---
        \textbf{Model Prediction:}
        \begin{tcolorbox}[
            breakable,
            colback=black!5!white,
            colframe=gray!20,
            boxrule=0.5pt,
            arc=0pt,
            left=1mm, right=1mm, top=1mm, bottom=1mm,
            fontupper=\small
        ]
        The speaker's tone of voice conveys a playful and affectionate sentiment. The use of \"I'll marry you\" suggests a light-hearted, teasing manner rather than a serious proposal. This is further emphasized by the bird sound effect in the background, which adds to the whimsical nature of the statement.
        \end{tcolorbox}
        
        \vspace{0.5em}
        
        % --- Part 3: Metric / Score ---
        \textbf{Score:} \hlToken{5.0}\\
        
        % --- Part 3: Evaluation & Score ---
        \textbf{Judge Response:}
        \begin{tcolorbox}[
            breakable,
            colback=blue!5!white,
            colframe=blue!15!gray,
            boxrule=0.5pt,
            arc=0pt,
            left=1mm, right=1mm, top=1mm, bottom=1mm,
            fontupper=\small
        ]
        Explanation: The reference answer indicates that the speaker's tone suggests a positive sentiment. The model's answer elaborates on this by describing the tone as playful and affectionate, which aligns with a positive sentiment. It also provides context for the tone by mentioning the light-heartedness of the phrase \"I'll marry you\" and the whimsical bird sound effect. This adds depth to the interpretation while remaining relevant to the overall positive sentiment described in the reference answer. Therefore, the model's response accurately captures the essence of the reference.
        \end{tcolorbox}

        \textit{Note: The bird sound effect in this case was an artificially superimposed real-world background interference. The model's response indicates that ALLMs tend to assign emotional connotations to specific environmental noises (e.g., interpreting bird chirps as "whimsical"). While the judgment happened to be correct here, it vividly illustrates how real-world interference can bias the model’s extraction and interpretation of affective cues.}
        
    \end{tcolorbox}
\end{figure*}

\begin{figure*}[t!]
    \centering
    \begin{tcolorbox}[
        %breakable,
        colback=gray!5!white,
        colframe=ERcolor,
        title={ER (LLM-as-a-Judge Metric)},
        fonttitle=\bfseries,
        sharp corners,
        left=2mm, right=2mm, top=2mm, bottom=2mm
     ]
        % --- Part 1: Question & Reference ---
        \textbf{Instruction:} How would you interpret the speaker's sentiment state from their speech?\\
        \vspace{0.3em}
        
        \textbf{Reference:}
        \begin{tcolorbox}[
            colback=black!5!white,
            colframe=gray!20,
            boxrule=0.5pt,
            arc=0pt,
            left=1mm, right=1mm, top=1mm, bottom=1mm,
            fontupper=\small
        ]
        The speaker's tone suggesting a negative sentiment.
        \end{tcolorbox}
        \vspace{0.5em}

        % --- Part 2: Model Prediction ---
        \textbf{Model Prediction:}
        \begin{tcolorbox}[
            breakable,
            colback=black!5!white,
            colframe=gray!20,
            boxrule=0.5pt,
            arc=0pt,
            left=1mm, right=1mm, top=1mm, bottom=1mm,
            fontupper=\small
        ]
        The speaker's tone of voice conveys excitement and urgency, as indicated by the exclamation 'get some ice here!'
        \end{tcolorbox}
        
        \vspace{0.5em}
        
        % --- Part 3: Metric / Score ---
        \textbf{Score:} \hlToken{0.0}\\
        
        % --- Part 3: Evaluation & Score ---
        \textbf{Judge Response:}
        \begin{tcolorbox}[
            breakable,
            colback=blue!5!white,
            colframe=blue!15!gray,
            boxrule=0.5pt,
            arc=0pt,
            left=1mm, right=1mm, top=1mm, bottom=1mm,
            fontupper=\small
        ]
        Explanation: The reference answer indicates a negative sentiment in the speaker's tone, while the model's answer describes a tone of excitement and urgency, which is the opposite of what the reference suggests. The model's response does not align with the sentiment presented in the reference, providing incorrect information. Therefore, the model's answer is not accurate or relevant in this context.
        \end{tcolorbox}
    \end{tcolorbox}
\end{figure*}

\begin{figure*}[t!]
    \centering
    \begin{tcolorbox}[
        %breakable, 
        colback=gray!5!white,
        colframe=MRcolor,
        title={MR (Acc Metric)},
        fonttitle=\bfseries,
        sharp corners,
        left=2mm, right=2mm, top=2mm, bottom=2mm
     ]
        \textbf{Instruction:} Carefully listen to the equation and calculate the final value.\\
        \vspace{0.3em}

        \textbf{Audio Content:} what’s 33.4 minus 32.56?\\
        \vspace{0.3em}

        \textbf{Reference:}
        \begin{tcolorbox}[
            colback=black!5!white,
            colframe=gray!20,
            boxrule=0.5pt,
            arc=0pt,
            left=1mm, right=1mm, top=1mm, bottom=1mm,
            fontupper=\small
        ]
        0.84
        \end{tcolorbox}
        \vspace{0.5em}

        % --- Part 2: Model Prediction ---
        \textbf{Model Prediction:}
        \begin{tcolorbox}[
            colback=black!5!white,
            colframe=gray!20,
            boxrule=0.5pt,
            arc=0pt,
            left=1mm, right=1mm, top=1mm, bottom=1mm,
            fontupper=\small
        ]
        To find the result of 33.4 minus 32.56, follow these steps: 

        1. Subtract the numbers:
           - Line up the decimal points
           - Subtract each digit from right to left

        Here's a step-by-step breakdown:
        \begin{verbatim}
          33.40
        - 32.56
        -------
        \end{verbatim}

        2. Start subtracting from the rightmost digits:
           - The tenths place (0.4) is greater than the hundredths place (0.6), so no borrowing is needed.
           - 4 - 6 = -2, but since we can't have a negative number in this position, borrow 1 from the ones place.
        
        ... (omitted specific incorrect borrowing logic for brevity) ...

        So the final calculation results in:
        \begin{verbatim}
          0.83
        \end{verbatim}
        
        Therefore, 33.4 minus 32.56 equals 0.83.
        \end{tcolorbox}
        
        \vspace{0.5em}
        
        % --- Part 3: Metric / Score ---
        \textbf{Score:} \hlToken{0.0}\\
    \end{tcolorbox}
\end{figure*}

\begin{figure*}[t!]
    \centering
    \begin{tcolorbox}[
        %breakable, 
        colback=gray!5!white,
        colframe=MRcolor,
        title={MR (Acc Metric)},
        fonttitle=\bfseries,
        sharp corners,
        left=2mm, right=2mm, top=2mm, bottom=2mm
     ]
        \textbf{Instruction:} Extract the numbers and operation from the speech, solve the problem.\\
        \vspace{0.3em}

        \textbf{Audio Content:} calculate 660 minus 458?\\
        \vspace{0.3em}

        \textbf{Reference:}
        \begin{tcolorbox}[
            colback=black!5!white,
            colframe=gray!20,
            boxrule=0.5pt,
            arc=0pt,
            left=1mm, right=1mm, top=1mm, bottom=1mm,
            fontupper=\small
        ]
        202
        \end{tcolorbox}
        \vspace{0.5em}

        % --- Part 2: Model Prediction ---
        \textbf{Model Prediction:}
        \begin{tcolorbox}[
            colback=black!5!white,
            colframe=gray!20,
            boxrule=0.5pt,
            arc=0pt,
            left=1mm, right=1mm, top=1mm, bottom=1mm,
            fontupper=\small
        ]
        To solve this problem, we need to calculate the area of a circle with radius 6 cm and subtract it from the total surface area.

        1. Calculate the circumference:
           - Formula: \( C = 2\pi r \)
           - Given radius (\( r \)) = 6 cm
           - Circumference (\( C \)) = \( 2 \times \pi \times 6 = 37.68 \) cm
        
        2. Calculate the area of the circle:
           - Formula: \( A = \pi r^2 \)
           - Area (\( A \)) = \( \pi \times 6^2 = 113.04 \) cm\textsuperscript{2}
        
        3. Subtract the area of the circle from the total surface area (25 square decimeters):
           - Convert 25 dm\textsuperscript{2} to cm\textsuperscript{2}: \( 25 \times 100 = 2500 \) cm\textsuperscript{2}
           - Remaining area after removing the circular part: \( 2500 - 113.04 = 2386.96 \) cm\textsuperscript{2}
        
        Therefore, the remaining area is approximately 2386.96 cm\textsuperscript{2}.
        \end{tcolorbox}
        
        \vspace{0.5em}
        
        % --- Part 3: Metric / Score ---
        \textbf{Score:} \hlToken{0.0}\\
    \end{tcolorbox}
\end{figure*}

\begin{figure*}[t!]
    \centering
    \begin{tcolorbox}[
        %breakable, 
        colback=gray!5!white,
        colframe=SQAcolor,
        title={SQA (LLM-as-a-Judge Metric)},
        fonttitle=\bfseries,
        sharp corners,
        left=2mm, right=2mm, top=2mm, bottom=2mm
     ]
        \textbf{Question:} How long did it take for the Theses printing to spread through Germany?\\
        \vspace{0.3em}

        \textbf{Reference:}
        \begin{tcolorbox}[
            colback=black!5!white,
            colframe=gray!20,
            boxrule=0.5pt,
            arc=0pt,
            left=1mm, right=1mm, top=1mm, bottom=1mm,
            fontupper=\small
        ]
        two weeks
        \end{tcolorbox}
        \vspace{0.5em}

        % --- Part 2: Model Prediction ---
        \textbf{Model Prediction:}
        \begin{tcolorbox}[
            colback=black!5!white,
            colframe=gray!20,
            boxrule=0.5pt,
            arc=0pt,
            left=1mm, right=1mm, top=1mm, bottom=1mm,
            fontupper=\small
        ]
        The Theses of Martin Luther, which were printed and distributed in Wittenberg on October 31, 1517, quickly spread throughout Germany. Within a few months, they had reached many parts of the Holy Roman Empire, including major cities like Nuremberg and Augsburg. The rapid dissemination was facilitated by the advent of printing technology, particularly the invention of movable type by Johannes Gutenberg around that time. This allowed for more copies to be produced and circulated much faster than through handwritten manuscripts alone.The widespread distribution of these theses played a significant role in sparking the Protestant Reformation, as it brought about increased debate among scholars and clergy regarding issues such as papal authority, indulgences, and church practices.
        \end{tcolorbox}
        
        \vspace{0.5em}
        
        % --- Part 3: Metric / Score ---
        \textbf{Score:} \hlToken{1.0}\\

        \textbf{Judge Response:}
        \begin{tcolorbox}[
            breakable,
            colback=blue!5!white,
            colframe=blue!15!gray,
            boxrule=0.5pt,
            arc=0pt,
            left=1mm, right=1mm, top=1mm, bottom=1mm,
            fontupper=\small
        ]
        Explanation: The reference answer states that it took \"two weeks\" for the Theses to spread throughout Germany, while the model's answer provides a detailed explanation of the context and significance of the Theses' dissemination but does not mention the specific timeframe of two weeks. Therefore, the model's response diverges significantly from the reference in providing an accurate answer to the question asked. It does not address the duration explicitly, focusing instead on the impact and the means of dissemination. I think this results in a low alignment with the reference answer regarding the specific detail of the timeframe.
        \end{tcolorbox}
        
    \end{tcolorbox}
\end{figure*}

\begin{figure*}[t!]
    \centering
    \begin{tcolorbox}[
        %breakable, 
        colback=gray!5!white,
        colframe=SQAcolor,
        title={SQA (LLM-as-a-Judge Metric)},
        fonttitle=\bfseries,
        sharp corners,
        left=2mm, right=2mm, top=2mm, bottom=2mm
    ]
        \textbf{Question:} Whose English translation of the Bible did the Luther Bible influence?\\
        \vspace{0.3em}

        \textbf{Reference:}
        \begin{tcolorbox}[
            colback=black!5!white,
            colframe=gray!20,
            boxrule=0.5pt,
            arc=0pt,
            left=1mm, right=1mm, top=1mm, bottom=1mm,
            fontupper=\small
        ]
        william tyndale's
        \end{tcolorbox}
        \vspace{0.5em}

        % --- Part 2: Model Prediction ---
        \textbf{Model Prediction:}
        \begin{tcolorbox}[
            colback=black!5!white,
            colframe=gray!20,
            boxrule=0.5pt,
            arc=0pt,
            left=1mm, right=1mm, top=1mm, bottom=1mm,
            fontupper=\small
        ]
        The Luther Bible influenced the King James Version of the Bible.
        \end{tcolorbox}
        
        \vspace{0.5em}
        
        % --- Part 3: Metric / Score ---
        \textbf{Score:} \hlToken{0.0}\\

        \textbf{Judge Response:}
        \begin{tcolorbox}[
            breakable,
            colback=blue!5!white,
            colframe=blue!15!gray,
            boxrule=0.5pt,
            arc=0pt,
            left=1mm, right=1mm, top=1mm, bottom=1mm,
            fontupper=\small
        ]
        Explanation: The reference answer is focused on identifying William Tyndale as the translator whose work was influenced by the Luther Bible. In contrast, the model's answer discusses the impact of the Luther Bible on the King James Version, which does not directly address the question about Tyndale's translation. The model's response is somewhat relevant to the broader topic of English translations but fails to accurately answer the specific question regarding Tyndale. Therefore, it diverges significantly from the reference in terms of relevance and accuracy. I think the model answer is misaligned with the reference answer because it does not mention Tyndale or directly relate to the question asked. 
        \end{tcolorbox}
        
    \end{tcolorbox}
\end{figure*}

\begin{figure*}[t!]
    \centering
    \begin{tcolorbox}[
        %breakable, 
        colback=gray!5!white,
        colframe=SIcolor,
        title={SI (LLM-as-a-Judge Metric)},
        fonttitle=\bfseries,
        sharp corners,
        left=2mm, right=2mm, top=2mm, bottom=2mm
    ]

        \textbf{Instruction:} Please listen carefully and follow the instructions given in the audio.\\
        \vspace{0.3em}
    
        \textbf{Audio Content:} Consider a connected graph G with seven vertices and ten edges. Determine the genus of the graph.\\
        \vspace{0.3em}

        \textbf{Reference:}
        \begin{tcolorbox}[
            colback=black!5!white,
            colframe=gray!20,
            boxrule=0.5pt,
            arc=0pt,
            left=1mm, right=1mm, top=1mm, bottom=1mm,
            fontupper=\small
        ]
        The genus $g$ of a graph $G$ with $v$ vertices and $e$ edges can be determined using Euler's formula for a connected graph embedded on a surface of genus $g$: $v - e + f = 2 - 2g$, where $f$ is the number of faces in the embedding.

        For $G$ with $v = 7$ vertices and $e = 10$ edges, we solve for $f$:
        \[ 7 - 10 + f = 2 - 2g \]
        \[ f = 5 + 2g \]

        Substitute back into Euler's formula:
        \[ 7 - 10 + (5 + 2g) = 2 - 2g \]
        \[ 2 + 2g = 2 - 2g \]
        \[ 4g = 0 \]
        \[ g = 0 \]

        The genus of the graph $G$ is 0, meaning it can be embedded on a plane or a sphere.
        \end{tcolorbox}
        \vspace{0.5em}

        % --- Part 2: Model Prediction ---
        \textbf{Model Prediction:}
        \begin{tcolorbox}[
            colback=black!5!white,
            colframe=gray!20,
            boxrule=0.5pt,
            arc=0pt,
            left=1mm, right=1mm, top=1mm, bottom=1mm,
            fontupper=\small
        ]
        A connected graph with 7 vertices and 10 edges is a tree. In graph theory, a tree is the simplest type of connected graph where there are no cycles. A tree with n vertices has exactly n-1 edges, so in this case, it's a tree because 7 - 1 = 6, which equals the number of edges (10).
        \end{tcolorbox}
        
        \vspace{0.5em}
        
        % --- Part 3: Metric / Score ---
        \textbf{Score:} \hlToken{1.0} \hfill \\

        \textbf{Judge Response:}
        \begin{tcolorbox}[
            breakable,
            colback=blue!5!white,
            colframe=blue!15!gray,
            boxrule=0.5pt,
            arc=0pt,
            left=1mm, right=1mm, top=1mm, bottom=1mm,
            fontupper=\small
        ]
        Explanation: The reference answer accurately applies Euler's formula to determine the genus of the graph, concluding that the genus is 0, indicating that the graph can be embedded on a plane or a sphere. In contrast, the model's answer incorrectly states that the graph is a tree, which is not possible given that a tree with 7 vertices would have only 6 edges, not 10. Thus, the model's answer provides incorrect information and does not address the genus calculation or the specific characteristics of the graph as outlined in the reference. Therefore, the model's answer is completely misaligned with the reference answer, as it fails to provide relevant and accurate information about the genus of the graph.
        \end{tcolorbox}
        
    \end{tcolorbox}
\end{figure*}

\begin{figure*}[t!]
    \centering
    \begin{tcolorbox}[
        %breakable, 
        colback=gray!5!white,
        colframe=SIcolor,
        title={SI (LLM-as-a-Judge Metric)},
        fonttitle=\bfseries,
        sharp corners,
        left=2mm, right=2mm, top=2mm, bottom=2mm
    ]

        \textbf{Instruction:} Please comply with the instructions given in the audio clip.\\
        \vspace{0.3em}
    
        \textbf{Audio Content:} How does a virus replicate and spread within its host cell, and what are the main mechanisms by which viruses are transmitted between hosts?\\
        \vspace{0.3em}

        \textbf{Reference:}
        \begin{tcolorbox}[
            colback=black!5!white,
            colframe=gray!20,
            boxrule=0.5pt,
            arc=0pt,
            left=1mm, right=1mm, top=1mm, bottom=1mm,
            fontupper=\small
        ]
        A virus replicates by injecting its genetic material into a host cell, hijacking the cell's machinery to produce more viral particles, which are then released to infect other cells. Viruses can spread between hosts through direct contact, respiratory droplets, contaminated surfaces, or vectors like insects.
        \end{tcolorbox}
        \vspace{0.5em}

        % --- Part 2: Model Prediction ---
        \textbf{Model Prediction:}
        \begin{tcolorbox}[
            colback=black!5!white,
            colframe=gray!20,
            boxrule=0.5pt,
            arc=0pt,
            left=1mm, right=1mm, top=1mm, bottom=1mm,
            fontupper=\small
        ]
        The virus replicates and spreads from host to host through several mechanisms: \\
        1. Direct Contact Transmission - The most common method, where the virus is transmitted directly between individuals via respiratory droplets or bodily fluids. This occurs when an infected person coughs, sneezes, talks, sings, or breathes near another person. \\
        2. Indirect Contact Transmission - When a person touches a surface contaminated with the virus (e.g., doorknobs, handrails) and then touches their face without washing their hands first. The virus can enter the body through mucous membranes in the eyes, nose, or mouth. \\
        3. Droplet Transmission - When an infected individual releases infectious particles into the air by talking, coughing, or sneezing. These droplets can land on surfaces or be inhaled by others nearby. \\
        4. Airborne Transmission - In some cases, viruses become aerosolized, meaning they are suspended in tiny airborne particles that can remain in the air for longer periods. This type of transmission is more likely under certain conditions like crowded indoor spaces with poor ventilation. \\
        \end{tcolorbox}
        
        \vspace{0.5em}
        
        % --- Part 3: Metric / Score ---
        \textbf{Score:} \hlToken{2.0} \hfill \\

        \textbf{Judge Response:}
        \begin{tcolorbox}[
            breakable,
            colback=blue!5!white,
            colframe=blue!15!gray,
            boxrule=0.5pt,
            arc=0pt,
            left=1mm, right=1mm, top=1mm, bottom=1mm,
            fontupper=\small
        ]
        Explanation: The reference answer provides a brief overview of how a virus replicates and spreads, focusing on the injection of genetic material into a host cell and the subsequent use of the cell's machinery to produce more viral particles. It also mentions various transmission methods. In contrast, the model answer elaborates on the transmission mechanisms in detail but does not address how a virus replicates within its host cell, which is a critical part of the question. Therefore, while the model answer is comprehensive regarding transmission methods, it lacks a complete response to the replication aspect mentioned in the reference. I think this discrepancy significantly affects the overall alignment with the reference answer.
        \end{tcolorbox}
        
    \end{tcolorbox}
\end{figure*}

\onecolumn

\clearpage  % 强制渲染图片并另起一页

\begin{multicols}{2} 
\section{Detailed Experimental Results}
\label{data}

In this section, we provide the complete quantitative results for all experiments conducted in this study. This includes the comprehensive performance of all evaluated ALLMs across six tasks and four acoustic ecologies, as well as the detailed ablation results for various denoising mitigation strategies.
\end{multicols}

\scriptsize
\setlength{\tabcolsep}{8.2pt}
\renewcommand{\arraystretch}{1.15}
\definecolor{mygrey}{HTML}{F5F5F5}
\definecolor{mygrey2}{HTML}{DCDCDC}
% ---- (可选) 如果你用到 \Xhline，请确保导言区有 makecell ----
% \usepackage{makecell}
% \usepackage{multirow}
% \usepackage[table]{xcolor}

% ====== 两个小宏：Scene 标题行 + 任务对表头（Model 垂直居中） ======

\newcommand{\SceneHeader}[1]{%
  % rule[-深度]{0pt}{总高度}：-1.5ex 确保文字下方有留白，4.5ex 确保上方有留白
  \rowcolor{mygrey2}\multicolumn{11}{l}{\rule[-1.5ex]{0pt}{4.5ex}Scene: #1}\\
  \Xhline{1.2pt}
}

\newcommand{\TaskPairHeader}[2]{%
  \rowcolor{mygrey}
  % \multirow{2}{*}[0.5ex]：这里的 [0.5ex] 是关键，它将文字微调向上，防止被第二行的颜色压住
  \multirow{2}{*}[0.5ex]{Model} & 
  \multicolumn{5}{c|}{\rule{0pt}{3.5ex}#1} & 
  \multicolumn{5}{c}{#2} \\
  \rowcolor{mygrey}
  % 为第二行表头增加支柱，确保 K=0 等文字上下都有空间
  \rule[-1.2ex]{0pt}{3.2ex} & $K=0$ & $K=1$ & $K=2$ & $K=3$ & $K=4$
  & $K=0$ & $K=1$ & $K=2$ & $K=3$ & $K=4$ \\
  \Xhline{0.9pt}
}

\begin{longtable}{lccccc|ccccc}

\label{tab:graph3_all_4d}\\

\endfirsthead
\endhead

% ===================== 页脚（用薄线，避免翻页处“黑线太重”） =====================
\Xhline{1.2pt}
\multicolumn{11}{r}{\scriptsize\textit{Continued on next page}}\\
\endfoot

% ===================== 最后一页页脚（只保留这一条最终粗线） =====================
\Xhline{2pt}
\endlastfoot

\Xhline{1.2pt}
\SceneHeader{Pasture}
\TaskPairHeader{ASR}{MR}
\textit{Qwen2-Audio} & 3.45 & 4.67 & 8.08 & 14.48 & 24.80 & 66.00 & 58.00 & 39.00 & 20.00 & 10.00 \\
\textit{SALMONN} & 10.49 & 16.16 & 28.38 & 73.41 & 161.12 & 18.00 & 8.00 & 5.00 & 3.00 & 0.00 \\
\textit{SeaLLMs} & 5.52 & 19.05 & 15.78 & 36.55 & 65.47 & 62.00 & 52.00 & 31.00 & 22.00 & 9.00 \\
\textit{}Phi-4 & 1.67 & 2.65 & 5.35 & 12.80 & 25.97 & 3.00 & 5.00 & 1.00 & 1.00 & 0.00 \\
\textit{MERaLION} & 2.34 & 5.32 & 12.74 & 22.49 & 38.13 & 74.00 & 67.00 & 48.00 & 28.00 & 15.00 \\
\textit{StepAudio2} & 3.90 & 5.27 & 7.81 & 14.10 & 25.94 & 75.00 & 66.00 & 48.00 & 30.00 & 14.00 \\
\textit{MiniCPM} & 2.95 & 7.18 & 18.64 & 38.56 & 68.03 & 75.00 & 65.00 & 38.00 & 19.00 & 7.00 \\
\textit{Qwen-Turbo} & 23.78 & 25.10 & 30.47 & 40.69 & 55.09 & 88.00 & 66.00 & 33.00 & 20.00 & 14.00 \\
\textit{Qwen2.5-Omni} & 23.32 & 25.71 & 30.09 & 43.17 & 56.07 & 89.00 & 58.00 & 37.00 & 23.00 & 14.00 \\
\textit{Qwen3-Omni} & 1.72 & 2.25 & 5.12 & 10.61 & 21.96 & 91.00 & 86.00 & 64.00 & 44.00 & 25.00 \\
\textit{GPT-4o-Audio} & 50.01 & 60.46 & 72.85 & 87.32 & 100.04 & 93.00 & 74.00 & 50.00 & 29.00 & 15.00 \\
\Xhline{1.1pt}

\TaskPairHeader{ER}{SQA}
\textit{Qwen2-Audio} & 51.53 & 35.25 & 31.61 & 27.62 & 26.25 & 79.85 & 78.63 & 76.03 & 75.25 & 73.38 \\
\textit{SALMONN} & 40.53 & 27.61 & 25.63 & 25.13 & 25.13 & 79.90 & 75.69 & 70.39 & 68.92 & 65.34 \\
\textit{SeaLLMs} & 47.80 & 21.64 & 21.11 & 18.65 & 17.16 & 78.58 & 76.86 & 75.05 & 70.59 & 65.44 \\
\textit{Phi-4} & 49.92 & 24.29 & 23.06 & 20.19 & 18.23 & 85.74 & 84.36 & 83.97 & 82.79 & 81.08 \\
\textit{MERaLION} & 52.60 & 53.90 & 52.26 & 46.59 & 43.25 & 80.69 & 81.27 & 81.23 & 73.68 & 78.92 \\
\textit{StepAudio2} & 56.81 & 38.00 & 35.05 & 31.68 & 30.07 & 81.37 & 80.54 & 77.01 & 75.69 & 72.94 \\
\textit{MiniCPM} & 55.32 & 28.35 & 27.81 & 26.66 & 26.24 & 82.94 & 83.87 & 82.35 & 79.75 & 76.96 \\
\textit{Qwen-Turbo} & 52.99 & 23.68 & 23.56 & 19.96 & 15.63 & 82.50 & 84.17 & 83.33 & 79.31 & 75.00 \\
\textit{Qwen2.5-Omni } & 52.91 & 24.10 & 24.06 & 19.50 & 18.47 & 83.82 & 84.90 & 82.16 & 79.02 & 75.88 \\
\textit{Qwen3-Omni} & 47.20 & 32.03 & 29.81 & 28.47 & 26.55 & 80.98 & 81.52 & 82.50 & 81.76 & 80.10 \\
\textit{GPT-4o-Audio} & 30.61 & 5.90 & 5.17 & 1.88 & 0.92 & 86.62 & 85.83 & 86.37 & 86.32 & 85.98 \\

\Xhline{1.1pt}

\TaskPairHeader{GR}{SI}
\textit{Qwen2-Audio} & 96.02 & 94.52 & 94.82 & 95.12 & 94.12 & 49.60 & 46.60 & 45.80 & 35.20 & 30.40 \\
\textit{SALMONN} & 82.37 & 84.86 & 81.87 & 76.29 & 76.29 & 58.40 & 53.00 & 55.00 & 53.20 & 56.00 \\
\textit{SeaLLMs} & 79.87 & 72.73 & 76.01 & 69.62 & 72.62 & 62.00 & 41.40 & 37.60 & 26.60 & 20.20 \\
\textit{Phi-4} & 38.65 & 43.13 & 36.35 & 31.47 & 24.20 & 33.20 & 17.40 & 20.00 & 22.40 & 18.00 \\
\textit{MERaLION} & 85.26 & 84.36 & 83.86 & 82.47 & 78.98 & 71.00 & 68.80 & 62.40 & 55.00 & 42.60 \\
\textit{StepAudio2} & 86.95 & 88.45 & 83.37 & 83.76 & 77.19 & 58.20 & 50.80 & 44.80 & 40.80 & 35.20 \\
\textit{MiniCPM} & 93.43 & 90.04 & 90.24 & 87.75 & 84.66 & 72.40 & 71.80 & 66.20 & 58.60 & 41.00 \\
\textit{Qwen-Turbo} & 91.63 & 93.92 & 92.13 & 92.43 & 91.14 & 78.20 & 69.00 & 66.80 & 56.00 & 43.00 \\
\textit{Qwen2.5-Omni } & 91.53 & 92.03 & 92.83 & 91.43 & 90.34 & 76.60 & 71.20 & 67.80 & 53.80 & 45.20 \\
\textit{Qwen3-Omni} & 95.92 & 96.31 & 95.52 & 95.22 & 93.43 & 82.60 & 69.80 & 69.20 & 62.20 & 46.60 \\
\textit{GPT-4o-Audio} & \textcolor{gray}{--} & \textcolor{gray}{--} & \textcolor{gray}{--} & \textcolor{gray}{--} & \textcolor{gray}{--} & 78.20 & 79.20 & 68.60 & 59.40 & 38.20 \\
\Xhline{1.1pt}

\SceneHeader{Extreme Weather}
\TaskPairHeader{ASR}{MR}
\textit{Qwen2-Audio} & 3.45 & 6.90 & 14.06 & 24.64 & 38.99 & 66.00 & 51.00 & 29.00 & 15.00 & 3.00 \\
\textit{SALMONN} & 10.49 & 21.85 & 67.26 & 170.19 & 346.83 & 18.00 & 3.00 & 0.00 & 0.00 & 0.00 \\
\textit{SeaLLMs} & 5.52 & 16.84 & 44.96 & 64.75 & 142.99 & 62.00 & 41.00 & 17.00 & 8.00 & 5.00 \\
\textit{Phi-4} & 1.67 & 6.27 & 16.62 & 34.08 & 42.85 & 3.00 & 3.00 & 2.00 & 1.00 & 1.00 \\
\textit{MERaLION} & 2.34 & 8.79 & 20.99 & 38.33 & 56.16 & 74.00 & 58.00 & 39.00 & 15.00 & 7.00 \\
\textit{StepAudio2} & 3.90 & 8.00 & 14.72 & 25.68 & 38.88 & 75.00 & 55.00 & 35.00 & 14.00 & 4.00 \\
\textit{MiniCPM} & 2.95 & 14.87 & 33.43 & 64.42 & 80.03 & 75.00 & 49.00 & 30.00 & 10.00 & 5.00 \\
\textit{Qwen-Turbo} & 23.78 & 25.09 & 41.42 & 46.51 & 66.00 & 88.00 & 59.00 & 26.00 & 13.00 & 7.00 \\
\textit{Qwen2.5-Omni } & 23.32 & 25.37 & 37.23 & 49.77 & 66.23 & 89.00 & 52.00 & 35.00 & 10.00 & 7.00 \\
\textit{Qwen3-Omni} & 1.72 & 5.31 & 12.29 & 43.18 & 94.80 & 91.00 & 75.00 & 51.00 & 29.00 & 14.00 \\
\textit{GPT-4o-Audio} & 50.01 & 56.61 & 71.00 & 85.35 & 102.00 & 93.00 & 67.00 & 43.00 & 24.00 & 10.00 \\
\Xhline{1.1pt}

\TaskPairHeader{ER}{SQA}
\textit{Qwen2-Audio} & 51.53 & 37.97 & 36.09 & 35.40 & 35.40 & 79.85 & 78.24 & 75.59 & 71.47 & 67.55 \\
\textit{SALMONN} & 40.53 & 31.04 & 29.80 & 29.80 & 29.73 & 79.90 & 72.70 & 70.34 & 68.04 & 64.71 \\
\textit{SeaLLMs} & 47.80 & 22.06 & 21.76 & 18.65 & 17.54 & 78.58 & 78.04 & 70.69 & 64.22 & 59.95 \\
\textit{Phi-4} & 49.92 & 23.71 & 24.17 & 19.57 & 17.78 & 85.74 & 85.15 & 81.72 & 80.54 & 77.60 \\
\textit{MERaLION} & 52.60 & 53.86 & 56.32 & 51.64 & 50.07 & 80.69 & 80.20 & 79.80 & 76.62 & 74.90 \\
\textit{StepAudio2} & 56.81 & 39.92 & 39.38 & 39.08 & 39.04 & 81.37 & 81.42 & 77.89 & 74.51 & 71.67 \\
\textit{MiniCPM} & 55.32 & 29.27 & 30.45 & 30.38 & 30.84 & 82.94 & 82.65 & 81.72 & 76.52 & 76.86 \\
\textit{Qwen-Turbo} & 52.99 & 23.87 & 24.56 & 17.82 & 12.49 & 82.50 & 82.30 & 80.78 & 76.32 & 74.46 \\
\textit{Qwen2.5-Omni } & 52.91 & 23.45 & 25.56 & 17.66 & 14.75 & 83.82 & 83.04 & 81.27 & 80.10 & 75.59 \\
\textit{Qwen3-Omni} & 47.20 & 33.98 & 33.98 & 35.40 & 36.40 & 80.98 & 81.37 & 78.33 & 81.47 & 78.87 \\
\textit{GPT-4o-Audio} & 30.61 & 7.05 & 9.20 & 3.83 & 1.95 & 86.62 & 86.08 & 85.39 & 84.85 & 83.48 \\
\Xhline{1.1pt}

\TaskPairHeader{GR}{SI}
\textit{Qwen2-Audio} & 96.02 & 95.02 & 92.83 & 92.93 & 92.93 & 49.60 & 45.40 & 31.20 & 29.00 & 13.40 \\
\textit{SALMONN} & 82.37 & 83.27 & 75.40 & 69.22 & 64.04 & 58.40 & 53.00 & 51.80 & 54.20 & 56.80 \\
\textit{SeaLLMs} & 79.87 & 69.34 & 73.99 & 69.42 & 66.73 & 62.00 & 46.40 & 34.40 & 23.40 & 8.60 \\
\textit{Phi-4} & 38.65 & 40.14 & 39.74 & 30.08 & 22.21 & 33.20 & 23.60 & 19.40 & 20.60 & 13.00 \\
\textit{MERaLION} & 85.26 & 83.86 & 83.07 & 84.16 & 79.78 & 71.00 & 64.60 & 56.20 & 45.60 & 27.00 \\
\textit{StepAudio2} & 86.95 & 89.34 & 85.26 & 83.76 & 79.38 & 58.20 & 49.60 & 40.80 & 37.80 & 23.40 \\
\textit{MiniCPM} & 93.43 & 92.53 & 92.53 & 90.84 & 90.24 & 72.40 & 65.00 & 59.00 & 39.00 & 27.40 \\
\textit{Qwen-Turbo} & 91.63 & 94.72 & 92.93 & 92.83 & 92.13 & 78.20 & 73.00 & 62.80 & 49.80 & 36.60 \\
\textit{Qwen2.5-Omni } & 91.53 & 93.82 & 95.02 & 92.33 & 91.93 & 76.60 & 72.60 & 59.60 & 48.40 & 38.20 \\
\textit{Qwen3-Omni} & 95.92 & 95.42 & 95.62 & 95.22 & 94.22 & 82.60 & 75.60 & 62.60 & 57.40 & 41.80 \\
\textit{GPT-4o-Audio} & \textcolor{gray}{--} & \textcolor{gray}{--} & \textcolor{gray}{--} & \textcolor{gray}{--} & \textcolor{gray}{--} & 78.20 & 73.80 & 63.00 & 51.80 & 31.20 \\
\Xhline{1.1pt}

\SceneHeader{Classroom}
\TaskPairHeader{ASR}{MR}
\textit{Qwen2-Audio} & 3.45 & 4.24 & 6.47 & 9.95 & 14.63 & 66.00 & 62.00 & 46.00 & 33.00 & 26.00 \\
\textit{SALMONN} & 10.49 & 11.74 & 16.85 & 27.14 & 35.53 & 18.00 & 11.00 & 6.00 & 4.00 & 3.00 \\
\textit{SeaLLMs} & 5.52 & 7.42 & 15.08 & 27.63 & 52.31 & 62.00 & 57.00 & 47.00 & 35.00 & 27.00 \\
\textit{Phi-4} & 1.67 & 2.32 & 3.67 & 8.81 & 12.67 & 3.00 & 5.00 & 5.00 & 3.00 & 3.00 \\
\textit{MERaLION} & 2.34 & 3.46 & 6.10 & 10.34 & 15.70 & 74.00 & 71.00 & 69.00 & 52.00 & 41.00 \\
\textit{StepAudio2} & 3.90 & 5.67 & 7.56 & 8.66 & 11.68 & 75.00 & 70.00 & 61.00 & 50.00 & 38.00 \\
\textit{MiniCPM} & 2.95 & 6.25 & 10.36 & 18.39 & 29.66 & 75.00 & 73.00 & 56.00 & 48.00 & 38.00 \\
\textit{Qwen-Turbo} & 23.78 & 23.92 & 25.91 & 26.32 & 28.95 & 88.00 & 72.00 & 59.00 & 47.00 & 36.00 \\
\textit{Qwen2.5-Omni} & 23.32 & 24.16 & 24.32 & 26.06 & 28.83 & 89.00 & 69.00 & 55.00 & 46.00 & 36.00 \\
\textit{Qwen3-Omni} & 1.72 & 2.51 & 3.44 & 4.77 & 7.67 & 91.00 & 82.00 & 80.00 & 72.00 & 59.00 \\
\textit{GPT-4o-Audio} & 50.01 & 55.12 & 60.48 & 66.27 & 75.78 & 93.00 & 83.00 & 67.00 & 51.00 & 43.00 \\
\Xhline{1.1pt}

\TaskPairHeader{ER}{SQA}
\textit{Qwen2-Audio} & 51.53 & 37.51 & 36.05 & 35.78 & 35.01 & 79.85 & 78.68 & 78.28 & 75.54 & 72.45 \\
\textit{SALMONN} & 40.53 & 32.07 & 30.99 & 29.77 & 29.08 & 79.90 & 78.38 & 75.20 & 72.94 & 73.48 \\
\textit{SeaLLMs} & 47.80 & 22.75 & 22.83 & 19.88 & 19.11 & 78.58 & 77.21 & 77.75 & 75.88 & 73.77 \\
\textit{Phi-4} & 49.92 & 24.21 & 23.25 & 20.95 & 17.50 & 85.74 & 84.56 & 83.14 & 83.68 & 84.17 \\
\textit{MERaLION} & 52.60 & 54.98 & 55.93 & 52.87 & 49.96 & 80.69 & 81.76 & 79.75 & 81.76 & 79.61 \\
\textit{StepAudio2} & 56.81 & 39.11 & 38.31 & 37.31 & 33.86 & 81.37 & 81.76 & 79.56 & 77.21 & 76.91 \\
\textit{MiniCPM} & 55.32 & 30.26 & 28.96 & 27.81 & 27.24 & 82.94 & 83.19 & 82.60 & 83.58 & 81.32 \\
\textit{Qwen-Turbo} & 52.99 & 26.21 & 26.63 & 22.99 & 21.34 & 82.50 & 84.46 & 83.82 & 83.97 & 82.25 \\
\textit{Qwen2.5-Omni } & 52.91 & 26.93 & 27.43 & 23.37 & 22.18 & 83.82 & 84.71 & 83.63 & 83.68 & 81.27 \\
\textit{Qwen3-Omni} & 47.20 & 33.49 & 32.26 & 32.80 & 32.80 & 80.98 & 80.93 & 81.72 & 82.30 & 79.71 \\
\textit{GPT-4o-Audio} & 30.61 & 8.51 & 8.74 & 4.71 & 3.26 & 86.62 & 86.27 & 85.29 & 86.13 & 85.69 \\
\Xhline{1.1pt}

\TaskPairHeader{GR}{SI}
\textit{Qwen2-Audio} & 96.02 & 95.22 & 95.62 & 95.52 & 94.62 & 49.60 & 47.20 & 45.60 & 35.00 & 29.60 \\
\textit{SALMONN} & 82.37 & 78.49 & 76.20 & 72.81 & 64.54 & 58.40 & 57.40 & 54.00 & 54.80 & 56.20 \\
\textit{SeaLLMs} & 79.87 & 77.17 & 75.81 & 77.06 & 76.47 & 62.00 & 49.20 & 47.20 & 40.20 & 36.00 \\
\textit{Phi-4} & 38.65 & 39.64 & 38.84 & 33.96 & 29.88 & 33.20 & 21.40 & 20.00 & 20.40 & 19.60 \\
\textit{MERaLION} & 85.26 & 84.06 & 84.96 & 84.06 & 82.17 & 71.00 & 69.60 & 67.20 & 65.80 & 65.00 \\
\textit{StepAudio2} & 86.95 & 89.74 & 89.24 & 89.74 & 89.84 & 58.20 & 53.40 & 46.60 & 40.00 & 34.80 \\
\textit{MiniCPM} & 93.43 & 92.93 & 90.14 & 89.14 & 85.86 & 72.40 & 72.00 & 67.80 & 67.80 & 63.00 \\
\textit{Qwen-Turbo} & 91.63 & 94.02 & 94.02 & 94.32 & 94.22 & 78.20 & 73.60 & 76.00 & 69.40 & 64.80 \\
\textit{Qwen2.5-Omni } & 91.53 & 94.52 & 94.62 & 95.22 & 93.13 & 76.60 & 71.80 & 70.80 & 72.80 & 66.80 \\
\textit{Qwen3-Omni} & 95.92 & 96.31 & 96.12 & 95.92 & 94.12 & 82.60 & 74.20 & 72.80 & 77.00 & 73.40 \\
\textit{GPT-4o-Audio} & \textcolor{gray}{--} & \textcolor{gray}{--} & \textcolor{gray}{--} & \textcolor{gray}{--} & \textcolor{gray}{--} & 78.20 & 79.00 & 80.00 & 75.00 & 61.00 \\
\Xhline{1.1pt}

\SceneHeader{Outdoors}
\TaskPairHeader{ASR}{MR}
\textit{Qwen2-Audio} & 3.45 & 8.49 & 19.67 & 35.97 & 54.75 & 66.00 & 43.00 & 23.00 & 6.00 & 4.00 \\
\textit{SALMONN} & 10.49 & 24.47 & 124.79 & 317.33 & 509.11 & 18.00 & 5.00 & 2.00 & 0.00 & 0.00 \\
\textit{SeaLLMs} & 5.52 & 25.49 & 51.13 & 125.50 & 279.27 & 62.00 & 29.00 & 10.00 & 1.00 & 0.00 \\
\textit{Phi-4} & 1.67 & 7.07 & 19.54 & 42.89 & 81.12 & 3.00 & 1.00 & 2.00 & 2.00 & 1.00 \\
\textit{MERaLION} & 2.34 & 11.63 & 30.89 & 55.35 & 76.04 & 74.00 & 46.00 & 27.00 & 7.00 & 3.00 \\
\textit{StepAudio2} & 3.90 & 7.59 & 20.47 & 34.49 & 66.67 & 75.00 & 47.00 & 26.00 & 12.00 & 6.00 \\
\textit{MiniCPM} & 2.95 & 21.08 & 57.34 & 89.09 & 121.17 & 75.00 & 42.00 & 18.00 & 7.00 & 1.00 \\
\textit{Qwen-Turbo} & 23.78 & 27.95 & 42.30 & 61.54 & 96.42 & 88.00 & 48.00 & 18.00 & 7.00 & 4.00 \\
\textit{Qwen2.5-Omni} & 23.32 & 28.89 & 45.18 & 61.57 & 93.27 & 89.00 & 39.00 & 16.00 & 4.00 & 4.00 \\
\textit{Qwen3-Omni} & 1.72 & 5.70 & 48.41 & 259.56 & 557.20 & 91.00 & 63.00 & 40.00 & 18.00 & 5.00 \\
\textit{GPT-4o-Audio} & 50.01 & 64.69 & 86.97 & 107.27 & 118.39 & 93.00 & 49.00 & 16.00 & 6.00 & 3.00 \\
\Xhline{1.1pt}

\TaskPairHeader{ER}{SQA}
\textit{Qwen2-Audio} & 51.53 & 35.29 & 35.24 & 30.57 & 29.46 & 79.85 & 77.21 & 73.97 & 67.21 & 62.25 \\
\textit{SALMONN} & 40.53 & 30.87 & 30.45 & 30.45 & 30.22 & 79.90 & 73.14 & 67.84 & 63.82 & 62.75 \\
\textit{SeaLLMs} & 47.80 & 20.91 & 22.37 & 15.63 & 15.05 & 78.58 & 73.92 & 65.93 & 58.38 & 55.74 \\
\textit{Phi-4} & 49.92 & 22.79 & 23.86 & 18.16 & 13.90 & 85.74 & 86.42 & 80.69 & 80.05 & 73.28 \\
\textit{MERaLION} & 52.60 & 52.56 & 55.63 & 46.51 & 45.40 & 80.69 & 80.29 & 77.60 & 73.68 & 71.08 \\
\textit{StepAudio2} & 56.81 & 38.69 & 38.54 & 36.74 & 36.78 & 81.37 & 79.31 & 76.47 & 69.07 & 61.27 \\
\textit{MiniCPM} & 55.32 & 30.03 & 29.34 & 29.84 & 29.42 & 82.94 & 82.45 & 80.00 & 75.78 & 71.37 \\
\textit{Qwen-Turbo} & 52.99 & 22.91 & 25.02 & 14.10 & 10.57 & 82.50 & 82.65 & 79.71 & 70.20 & 66.62 \\
\textit{Qwen2.5-Omni} & 52.91 & 23.18 & 24.90 & 13.60 & 12.34 & 83.82 & 81.37 & 78.97 & 73.53 & 66.96 \\
\textit{Qwen3-Omni} & 47.20 & 34.10 & 33.56 & 33.41 & 34.10 & 80.98 & 80.83 & 81.03 & 75.74 & 73.53 \\
\textit{GPT-4o-Audio} & 30.61 & 5.67 & 8.93 & 1.07 & 0.23 & 86.62 & 86.18 & 86.18 & 83.38 & 79.17 \\

\Xhline{1.1pt}

\TaskPairHeader{GR}{SI}
\textit{Qwen2-Audio} & 96.02 & 93.63 & 90.04 & 87.65 & 81.77 & 49.60 & 43.20 & 32.60 & 19.80 & 6.60 \\
\textit{SALMONN} & 82.37 & 82.67 & 71.41 & 65.84 & 59.66 & 58.40 & 51.60 & 55.20 & 57.40 & 54.60 \\
\textit{SeaLLMs} & 79.87 & 71.04 & 74.19 & 73.24 & 75.66 & 62.00 & 39.20 & 21.60 & 10.00 & 3.60 \\
\textit{Phi-4} & 38.65 & 38.65 & 32.07 & 27.69 & 21.41 & 33.20 & 26.40 & 18.40 & 17.60 & 11.80 \\
\textit{MERaLION} & 85.26 & 82.97 & 84.06 & 81.77 & 76.49 & 71.00 & 67.40 & 55.20 & 37.40 & 17.20 \\
\textit{StepAudio2} & 86.95 & 85.96 & 81.67 & 83.76 & 77.19 & 58.20 & 51.60 & 36.40 & 23.80 & 9.80 \\
\textit{MiniCPM} & 93.43 & 91.33 & 90.14 & 85.16 & 81.87 & 72.40 & 59.80 & 46.80 & 30.20 & 12.00 \\
\textit{Qwen-Turbo} & 91.63 & 91.53 & 89.84 & 88.55 & 86.35 & 78.20 & 70.40 & 53.80 & 36.20 & 21.00 \\
\textit{Qwen2.5-Omni } & 91.53 & 91.04 & 90.14 & 89.54 & 87.05 & 76.60 & 68.60 & 56.60 & 34.20 & 20.00 \\
\textit{Qwen3-Omni} & 95.92 & 92.53 & 91.33 & 88.35 & 88.94 & 82.60 & 70.20 & 58.40 & 41.40 & 17.80 \\
\textit{GPT-4o-Audio} & \textcolor{gray}{--} & \textcolor{gray}{--} & \textcolor{gray}{--} & \textcolor{gray}{--} & \textcolor{gray}{--} & 78.20 & 76.00 & 52.00 & 29.60 & 6.40 \\
\Xhline{1.1pt}
\caption{Comprehensive results across four dimensions.
Blocks are scenes; rows are models; columns are noise-source count $K=0\ldots4$ for each task.
Values are percentages (numbers $<1$ are multiplied by 100). ASR reports WER (lower is better); others are higher-is-better.}

\end{longtable}

\twocolumn
\clearpage

\begin{table*}[t]
  \centering
  \captionsetup{font=scriptsize, skip=1pt}
  \scriptsize
  \setlength{\tabcolsep}{3.0pt}
  \renewcommand{\arraystretch}{0.95}
  \vspace{-0.6em}
  
  \definecolor{color1}{HTML}{DCDCDC} % Scene 背景色
  \definecolor{color2}{HTML}{F5F5F5}  % 任务表头背景色
  
  \resizebox{\textwidth}{!}{%
  \begin{tabular}{lcccc|lcccc}
  \Xhline{2pt}
  \rowcolor{color1}
  \multicolumn{10}{c}{Scene: Pasture\rule[-0.9ex]{0pt}{3.0ex}}\\
  \Xhline{1.2pt}
  \rowcolor{color2}
  ASR & $K=1$ & $K=2$ & $K=3$ & $K=4$ & MR & $K=1$ & $K=2$ & $K=3$ & $K=4$ \\
  \Xhline{0.9pt}
  \textit{Noise}        & 4.67 & 8.08 & 14.48 & 24.80 &
  \textit{Noise}        & 58.00 & 39.00 & 20.00 & 10.00 \\
  
  \textit{NoiseReduce}  & 12.60 & 28.39 & 51.31 & 73.63 &
  \textit{NoiseReduce}  & 32.00 & 17.00 & 4.00 & 1.00 \\
  
  \textit{AudioDenoise} & 7.28 & 17.21 & 34.17 & 56.87 &
  \textit{AudioDenoise} & 45.00 & 30.00 & 11.00 & 3.00 \\
  
  \textit{PyRNNoise}    & 15.79 & 34.59 & 60.09 & 81.43 &
  \textit{PyRNNoise}    & 35.00 & 9.00 & 5.00 & 0.00 \\
  
  \textit{DeepFilterNet}& 9.62 & 24.28 & 41.81 & 60.68 &
  \textit{DeepFilterNet}& 50.00 & 25.00 & 7.00 & 3.00 \\
  \Xhline{1.1pt}
  \rowcolor{color2}
  
  ER & $K=1$ & $K=2$ & $K=3$ & $K=4$ & SQA & $K=1$ & $K=2$ & $K=3$ & $K=4$ \\
  \Xhline{0.9pt}
  \textit{Noise} & 34.21 & 30.96 & 29.69 & 28.43 & \textit{Noise} & 78.63 & 76.03 & 75.25 & 73.38 \\
  \textit{NoiseReduce} & 31.23 & 27.28 & 29.43 & 22.95 & \textit{NoiseReduce} & 76.18 & 73.33 & 68.43 & 61.37 \\
  \textit{AudioDenoise} & 30.96 & 28.66 & 28.24 & 20.31 & \textit{AudioDenoise} & 78.09 & 74.71 & 69.12 & 63.77 \\
  \textit{PyRNNoise} & 31.88 & 30.08 & 27.36 & 29.43 & \textit{PyRNNoise} & 74.71 & 70.05 & 63.38 & 57.89 \\
  \textit{DeepFilterNet} & 33.49 & 31.23 & 28.66 & 32.18 & \textit{DeepFilterNet} & 79.36 & 75.74 & 69.36 & 64.46 \\
  \Xhline{1.1pt}
  \rowcolor{color2}
  GR & $K=1$ & $K=2$ & $K=3$ & $K=4$ & SI & $K=1$ & $K=2$ & $K=3$ & $K=4$ \\
  \Xhline{0.9pt}
  \textit{Noise} & 94.52 & 94.82 & 95.12 & 94.12 & \textit{Noise} & 46.60 & 45.80 & 35.20 & 30.40 \\
  \textit{NoiseReduce} & 88.94 & 87.35 & 81.97 & 81.18 & \textit{NoiseReduce} & 37.60 & 22.80 & 16.80 & 4.60 \\
  \textit{AudioDenoise} & 95.72 & 93.43 & 91.33 & 88.94 & \textit{AudioDenoise} & 42.80 & 34.40 & 23.40 & 15.20 \\
  \textit{PyRNNoise} & 91.83 & 88.25 & 82.67 & 74.70 & \textit{PyRNNoise} & 38.60 & 26.40 & 12.40 & 2.60 \\
  \textit{DeepFilterNet} & 92.03 & 90.94 & 92.03 & 89.64 & \textit{DeepFilterNet} & 41.80 & 31.40 & 20.00 & 12.60 \\
  \Xhline{1.5pt}
  \rowcolor{color1}
  \multicolumn{10}{c}{Scene: Classroom\rule[-0.9ex]{0pt}{3.0ex}}\\
  \Xhline{1.2pt}
  \rowcolor{color2}
  ASR & $K=1$ & $K=2$ & $K=3$ & $K=4$ & MR & $K=1$ & $K=2$ & $K=3$ & $K=4$ \\
  \Xhline{0.9pt}
  \textit{Noise}        & 4.24 & 6.47 & 9.95 & 14.63 &
  \textit{Noise}        & 62.00 & 46.00 & 33.00 & 26.00 \\
  
  \textit{NoiseReduce}  & 12.90 & 24.16 & 38.56 & 55.74 &
  \textit{NoiseReduce}  & 29.00 & 18.00 & 9.00 & 4.00 \\
  
  \textit{AudioDenoise} & 5.62 & 10.88 & 18.67 & 30.61 &
  \textit{AudioDenoise} & 55.00 & 37.00 & 27.00 & 14.00 \\
  
  \textit{PyRNNoise}    & 7.73 & 15.08 & 24.62 & 36.48 &
  \textit{PyRNNoise}    & 46.00 & 33.00 & 21.00 & 16.00 \\
  
  \textit{DeepFilterNet}& 7.71 & 14.03 & 22.51 & 32.62 &
  \textit{DeepFilterNet}& 44.00 & 32.00 & 21.00 & 18.00 \\ 
  \Xhline{1.1pt}
  \rowcolor{color2}
  ER & $K=1$ & $K=2$ & $K=3$ & $K=4$ & SQA & $K=1$ & $K=2$ & $K=3$ & $K=4$ \\
  \Xhline{0.9pt}
  \textit{Noise} & 37.47 & 36.28 & 34.98 & 33.72 & \textit{Noise} & 78.68 & 78.28 & 75.54 & 72.45 \\
  \textit{NoiseReduce} & 32.11 & 29.16 & 35.21 & 25.79 & \textit{NoiseReduce} & 76.13 & 73.48 & 70.20 & 63.77 \\
  \textit{AudioDenoise} & 36.28 & 34.14 & 34.21 & 23.64 & \textit{AudioDenoise} & 78.87 & 75.20 & 70.74 & 65.20 \\
  \textit{PyRNNoise} & 32.80 & 29.35 & 35.75 & 31.11 & \textit{PyRNNoise} & 79.71 & 76.08 & 74.80 & 70.49 \\
  \textit{DeepFilterNet} & 34.56 & 32.34 & 34.83 & 32.64 & \textit{DeepFilterNet} & 81.32 & 77.01 & 73.53 & 72.70 \\
  \Xhline{1.1pt}
  \rowcolor{color2}
  GR & $K=1$ & $K=2$ & $K=3$ & $K=4$ & SI & $K=1$ & $K=2$ & $K=3$ & $K=4$ \\
  \Xhline{0.9pt}
  \textit{Noise} & 95.22 & 95.62 & 95.52 & 94.62 & \textit{Noise} & 47.20 & 45.60 & 35.00 & 29.60 \\
  \textit{NoiseReduce} & 88.55 & 84.56 & 83.47 & 79.68 & \textit{NoiseReduce} & 33.20 & 29.20 & 21.60 & 8.40 \\
  \textit{AudioDenoise} & 94.82 & 92.73 & 93.53 & 93.23 & \textit{AudioDenoise} & 42.40 & 47.40 & 33.20 & 22.60 \\
  \textit{PyRNNoise} & 92.63 & 91.43 & 92.13 & 90.94 & \textit{PyRNNoise} & 47.60 & 40.20 & 35.80 & 25.40 \\
  \textit{DeepFilterNet} & 92.23 & 92.83 & 93.43 & 92.23 & \textit{DeepFilterNet} & 43.00 & 41.60 & 35.60 & 27.80 \\ 
  \Xhline{2pt}
  \end{tabular}%
  }
  \caption{Denoising mitigation for Qwen2-Audio across two acoustic scenarios.
  The table reports Qwen2-Audio performance under increasing multi-source acoustic interference ($K=1\ldots4$) in Pasture and Classroom.
  Each task block compares the no-denoise baseline (Noise) with four denoising methods; ASR is WER (lower is better) and other tasks are higher-is-better.}
  \label{tab:mitigation_qwen2_pasture_classroom}
  \end{table*}

\begin{table*}[t]
  \centering
  \captionsetup{font=scriptsize, skip=1pt}
  \scriptsize
  \setlength{\tabcolsep}{3.0pt}
  \renewcommand{\arraystretch}{0.95}
  
  \definecolor{color1}{HTML}{DCDCDC} % Scene 背景色
  \definecolor{color2}{HTML}{F5F5F5}  % 任务表头背景色
  
  \resizebox{\textwidth}{!}{%
  \begin{tabular}{lcccc|lcccc}
  \Xhline{2pt}
  \rowcolor{color1}
  \multicolumn{10}{c}{Scene: Pasture\rule[-0.9ex]{0pt}{3.0ex}}\\
  \Xhline{1.2pt}
  \rowcolor{color2}
  ASR & $K=1$ & $K=2$ & $K=3$ & $K=4$ & MR & $K=1$ & $K=2$ & $K=3$ & $K=4$ \\
  \Xhline{0.9pt}
  \textit{Noise}        & 5.32 & 12.74 & 22.49 & 38.13 &
  \textit{Noise}        & 67.00 & 48.00 & 28.00 & 15.00 \\
  
  \textit{NoiseReduce}  & 13.34 & 30.96 & 54.26 & 73.05 &
  \textit{NoiseReduce}  & 40.00 & 20.00 & 6.00 & 3.00 \\
  
  \textit{AudioDenoise} & 8.27 & 20.36 & 39.06 & 61.20 &
  \textit{AudioDenoise} & 52.00 & 34.00 & 13.00 & 6.00 \\
  
  \textit{PyRNNoise}    & 12.58 & 31.89 & 56.58 & 78.60 &
  \textit{PyRNNoise}    & 38.00 & 21.00 & 5.00 & 1.00 \\
  
  \textit{DeepFilterNet}& 7.64 & 19.29 & 37.01 & 56.09 &
  \textit{DeepFilterNet}& 52.00 & 28.00 & 8.00 & 9.00 \\
  
  \Xhline{1.1pt}
  \rowcolor{color2}
  
  ER & $K=1$ & $K=2$ & $K=3$ & $K=4$ & SQA & $K=1$ & $K=2$ & $K=3$ & $K=4$ \\
  \Xhline{0.9pt}
  \textit{Noise}        & 56.21 & 53.33 & 50.46 & 48.74 &
  \textit{Noise}        & 81.27 & 81.23 & 73.68 & 78.92 \\
  
  \textit{NoiseReduce}  & 51.42 & 50.65 & 47.43 & 44.48 &
  \textit{NoiseReduce}  & 78.43 & 80.39 & 75.00 & 70.74 \\
  
  \textit{AudioDenoise} & 52.76 & 50.65 & 47.16 & 38.47 &
  \textit{AudioDenoise} & 80.98 & 79.71 & 77.75 & 71.47 \\
  
  \textit{PyRNNoise}    & 49.46 & 47.74 & 46.93 & 43.52 &
  \textit{PyRNNoise}    & 78.53 & 75.44 & 71.08 & 68.53 \\
  
  \textit{DeepFilterNet}& 47.93 & 49.31 & 46.55 & 47.32 &
  \textit{DeepFilterNet}& 81.13 & 76.86 & 77.16 & 71.91 \\
  \Xhline{1.1pt}
  \rowcolor{color2}
  
  GR & $K=1$ & $K=2$ & $K=3$ & $K=4$ & SI & $K=1$ & $K=2$ & $K=3$ & $K=4$ \\
  \Xhline{0.9pt}
  \textit{Noise}        & 84.36 & 83.86 & 82.47 & 78.98 &
  \textit{Noise}        & 68.80 & 62.40 & 55.00 & 42.60 \\
  
  \textit{NoiseReduce}  & 83.07 & 81.47 & 75.20 & 67.13 &
  \textit{NoiseReduce}  & 64.60 & 48.60 & 31.00 & 15.00 \\
  
  \textit{AudioDenoise} & 83.86 & 80.88 & 76.89 & 67.13 &
  \textit{AudioDenoise} & 69.20 & 59.40 & 44.60 & 24.40 \\
  
  \textit{PyRNNoise}    & 79.68 & 76.29 & 71.61 & 65.54 &
  \textit{PyRNNoise}    & 63.40 & 49.00 & 25.20 & 6.00 \\
  
  \textit{DeepFilterNet}& 79.28 & 77.59 & 76.69 & 76.00 &
  \textit{DeepFilterNet}& 66.40 & 60.80 & 39.00 & 24.80 \\
  \Xhline{1.5pt}
  \rowcolor{color1}
  \multicolumn{10}{c}{Scene: Classroom\rule[-0.9ex]{0pt}{3.0ex}}\\
  \Xhline{1.2pt}
  \rowcolor{color2}
  
  ASR & $K=1$ & $K=2$ & $K=3$ & $K=4$ & MR & $K=1$ & $K=2$ & $K=3$ & $K=4$ \\
  \Xhline{0.9pt}
  \textit{Noise}        & 3.46 & 6.10 & 10.34 & 15.70 &
  \textit{Noise}        & 71.00 & 69.00 & 52.00 & 41.00 \\
  
  \textit{NoiseReduce}  & 12.70 & 25.08 & 41.62 & 60.09 &
  \textit{NoiseReduce}  & 44.00 & 27.00 & 13.00 & 8.00 \\
  
  \textit{AudioDenoise} & 4.88 & 9.93 & 18.56 & 30.15 &
  \textit{AudioDenoise} & 67.00 & 53.00 & 29.00 & 23.00 \\
  
  \textit{PyRNNoise}    & 5.53 & 11.36 & 19.84 & 29.82 &
  \textit{PyRNNoise}    & 53.00 & 38.00 & 28.00 & 19.00 \\
  
  \textit{DeepFilterNet}& 5.77 & 10.50 & 16.98 & 25.00 &
  \textit{DeepFilterNet}& 60.00 & 44.00 & 28.00 & 19.00 \\\Xhline{1.1pt}
  \rowcolor{color2}
  ER & $K=1$ & $K=2$ & $K=3$ & $K=4$ & SQA & $K=1$ & $K=2$ & $K=3$ & $K=4$ \\
  \Xhline{0.9pt}
  \textit{Noise}        & 56.59 & 55.90 & 55.36 & 54.67 &
  \textit{Noise}        & 81.76 & 79.75 & 81.76 & 79.61 \\
  
  \textit{NoiseReduce}  & 51.15 & 51.30 & 52.34 & 45.36 &
  \textit{NoiseReduce}  & 80.29 & 79.41 & 78.53 & 74.51 \\
  
  \textit{AudioDenoise} & 52.87 & 54.37 & 53.10 & 45.33 &
  \textit{AudioDenoise} & 80.83 & 81.52 & 80.93 & 78.38 \\
  
  \textit{PyRNNoise}    & 51.84 & 51.19 & 51.92 & 49.35 &
  \textit{PyRNNoise}    & 80.78 & 80.39 & 79.41 & 78.28 \\
  
  \textit{DeepFilterNet}& 46.93 & 50.96 & 52.68 & 47.85 &
  \textit{DeepFilterNet}& 80.05 & 79.51 & 80.93 & 77.70 \\
  \Xhline{1.1pt}
  \rowcolor{color2}
  GR & $K=1$ & $K=2$ & $K=3$ & $K=4$ & SI & $K=1$ & $K=2$ & $K=3$ & $K=4$ \\
  \Xhline{0.9pt}
  \textit{Noise}        & 84.06 & 84.96 & 84.06 & 82.17 &
  \textit{Noise}        & 69.60 & 67.20 & 65.80 & 65.00 \\
  
  \textit{NoiseReduce}  & 84.76 & 80.48 & 78.78 & 74.40 &
  \textit{NoiseReduce}  & 60.80 & 53.60 & 38.20 & 22.60 \\
  
  \textit{AudioDenoise} & 84.06 & 83.37 & 78.78 & 74.70 &
  \textit{AudioDenoise} & 69.60 & 64.20 & 60.40 & 52.80 \\
  
  \textit{PyRNNoise}    & 78.49 & 79.08 & 80.78 & 78.39 &
  \textit{PyRNNoise}    & 68.00 & 66.00 & 60.40 & 52.60 \\
  
  \textit{DeepFilterNet}& 79.38 & 80.08 & 77.89 & 78.59 &
  \textit{DeepFilterNet}& 67.80 & 65.20 & 63.20 & 50.00 \\
  \Xhline{2pt}
  \end{tabular}%
  }
  \caption{Denoising mitigation for MERaLION across two acoustic scenarios.
  The table reports MERaLION performance under increasing multi-source acoustic interference ($K=1\ldots4$) in Pasture and Classroom.
  Each task block compares the no-denoise baseline (Noise) with four denoising methods; ASR is WER (lower is better) and other tasks are higher-is-better.}
  \label{tab:mitigation_meralion_pasture_classroom}
  \end{table*}

\begin{table*}[t]
  \centering
  \captionsetup{font=scriptsize, skip=1pt}
  \scriptsize
  \setlength{\tabcolsep}{3.0pt}
  \renewcommand{\arraystretch}{0.95}
  \vspace{-0.6em}
  
  \definecolor{color1}{HTML}{DCDCDC} % Scene 背景色
  \definecolor{color2}{HTML}{F5F5F5}  % 任务表头背景色
  
  \resizebox{\textwidth}{!}{%
  \begin{tabular}{lcccc|lcccc}
  \Xhline{2pt}
  \rowcolor{color1}
  \multicolumn{10}{c}{Scene: Pasture\rule[-0.9ex]{0pt}{3.0ex}}\\
  \Xhline{1.2pt}
  \rowcolor{color2}
  ASR & $K=1$ & $K=2$ & $K=3$ & $K=4$ & MR & $K=1$ & $K=2$ & $K=3$ & $K=4$ \\
  \Xhline{0.9pt}
  \textit{Noise} & 5.27 & 7.81 & 14.10 & 25.94 & \textit{Noise} & 66.00 & 48.00 & 30.00 & 14.00 \\
  \textit{NoiseReduce} & 10.57 & 25.63 & 69.87 & 183.00 & \textit{NoiseReduce} & 40.00 & 22.00 & 6.00 & 2.00 \\
  \textit{AudioDenoise} & 6.97 & 13.74 & 34.34 & 55.88 & \textit{AudioDenoise} & 54.00 & 40.00 & 14.00 & 6.00 \\
  \textit{PyRNNoise} & 14.69 & 37.58 & 73.92 & 105.67 & \textit{PyRNNoise} & 42.00 & 15.00 & 7.00 & 1.00 \\
  \textit{DeepFilterNet} & 10.06 & 20.20 & 46.42 & 86.65 & \textit{DeepFilterNet} & 52.00 & 31.00 & 10.00 & 6.00 \\
  \Xhline{1.1pt}
  \rowcolor{color2}
  ER & $K=1$ & $K=2$ & $K=3$ & $K=4$ & SQA & $K=1$ & $K=2$ & $K=3$ & $K=4$ \\
  \Xhline{0.9pt}
  \textit{Noise} & 37.70 & 35.75 & 34.94 & 32.80 & \textit{Noise} & 80.54 & 77.01 & 75.69 & 72.94 \\
  \textit{NoiseReduce} & 35.10 & 32.34 & 34.32 & 23.64 & \textit{NoiseReduce} & 79.90 & 75.34 & 67.75 & 58.43 \\
  \textit{AudioDenoise} & 36.97 & 34.44 & 31.65 & 26.74 & \textit{AudioDenoise} & 79.61 & 77.01 & 70.54 & 65.74 \\
  \textit{PyRNNoise} & 37.89 & 35.63 & 33.60 & 32.45 & \textit{PyRNNoise} & 79.12 & 74.71 & 69.85 & 61.42 \\
  \textit{DeepFilterNet} & 38.66 & 37.13 & 32.45 & 39.04 & \textit{DeepFilterNet} & 81.62 & 80.34 & 75.78 & 68.09 \\
  \Xhline{1.1pt}
  \rowcolor{color2}
  GR & $K=1$ & $K=2$ & $K=3$ & $K=4$ & SI & $K=1$ & $K=2$ & $K=3$ & $K=4$ \\
  \Xhline{0.9pt}
  \textit{Noise} & 88.45 & 83.37 & 83.76 & 77.19 & \textit{Noise} & 50.80 & 44.80 & 40.80 & 35.20 \\
  \textit{NoiseReduce} & 88.45 & 85.96 & 84.46 & 79.28 & \textit{NoiseReduce} & 46.80 & 37.00 & 17.20 & 6.40 \\
  \textit{AudioDenoise} & 86.25 & 86.35 & 83.96 & 78.39 & \textit{AudioDenoise} & 48.40 & 39.40 & 29.00 & 12.80 \\
  \textit{PyRNNoise} & 88.94 & 84.96 & 78.69 & 73.51 & \textit{PyRNNoise} & 48.40 & 25.60 & 13.40 & 2.80 \\
  \textit{DeepFilterNet} & 86.95 & 83.37 & 80.58 & 77.89 & \textit{DeepFilterNet} & 55.40 & 42.20 & 24.60 & 12.00 \\
  
  \Xhline{1.5pt}
  \rowcolor{color1}
  \multicolumn{10}{c}{Scene: Classroom\rule[-0.9ex]{0pt}{3.0ex}}\\
  \Xhline{1.2pt}
  \rowcolor{color2}
  ASR & $K=1$ & $K=2$ & $K=3$ & $K=4$ & MR & $K=1$ & $K=2$ & $K=3$ & $K=4$ \\
  \Xhline{0.9pt}
  \textit{Noise}        & 5.67 & 7.56 & 8.66 & 11.68 &
  \textit{Noise}        & 70.00 & 61.00 & 50.00 & 38.00 \\
  
  \textit{NoiseReduce}  & 12.19 & 21.02 & 41.22 & 86.09 &
  \textit{NoiseReduce}  & 33.00 & 24.00 & 11.00 & 4.00 \\
  
  \textit{AudioDenoise} & 5.66 & 9.58 & 17.09 & 35.65 &
  \textit{AudioDenoise} & 53.00 & 44.00 & 29.00 & 18.00 \\
  
  \textit{PyRNNoise}    & 8.21 & 10.69 & 20.78 & 30.57 &
  \textit{PyRNNoise}    & 55.00 & 39.00 & 32.00 & 19.00 \\
  
  \textit{DeepFilterNet}& 7.03 & 13.07 & 20.20 & 28.51 &
  \textit{DeepFilterNet}& 60.00 & 42.00 & 32.00 & 17.00 \\
  \Xhline{1.1pt}
  \rowcolor{color2}
  
  ER & $K=1$ & $K=2$ & $K=3$ & $K=4$ & SQA & $K=1$ & $K=2$ & $K=3$ & $K=4$ \\
  \Xhline{0.9pt}
  \textit{Noise}        & 40.61 & 38.70 & 38.89 & 37.78 &
  \textit{Noise}        & 81.76 & 79.56 & 77.21 & 76.91 \\
  
  \textit{NoiseReduce}  & 38.66 & 37.32 & 35.79 & 33.87 &
  \textit{NoiseReduce}  & 78.09 & 76.52 & 69.80 & 67.79 \\
  
  \textit{AudioDenoise} & 39.43 & 37.32 & 36.70 & 26.74 &
  \textit{AudioDenoise} & 80.83 & 77.06 & 75.69 & 68.63 \\
  
  \textit{PyRNNoise}    & 38.01 & 37.32 & 36.09 & 34.60 &
  \textit{PyRNNoise}    & 79.02 & 80.34 & 78.63 & 77.35 \\
  
  \textit{DeepFilterNet}& 38.54 & 38.54 & 37.09 & 37.89 &
  \textit{DeepFilterNet}& 80.59 & 81.23 & 81.27 & 77.16 \\
  \Xhline{1.1pt}
  \rowcolor{color2}
  
  GR & $K=1$ & $K=2$ & $K=3$ & $K=4$ & SI & $K=1$ & $K=2$ & $K=3$ & $K=4$ \\
  \Xhline{0.9pt}
  \textit{Noise}        & 89.74 & 89.24 & 89.74 & 89.84 &
  \textit{Noise}        & 53.40 & 46.60 & 40.00 & 34.80 \\
  
  \textit{NoiseReduce}  & 89.34 & 88.75 & 87.25 & 83.86 &
  \textit{NoiseReduce}  & 46.40 & 38.60 & 22.20 & 11.40 \\
  
  \textit{AudioDenoise} & 89.44 & 87.85 & 86.85 & 87.85 &
  \textit{AudioDenoise} & 54.00 & 46.60 & 39.20 & 29.00 \\
  
  \textit{PyRNNoise}    & 91.14 & 88.05 & 85.86 & 86.85 &
  \textit{PyRNNoise}    & 54.60 & 46.00 & 43.60 & 33.00 \\
  
  \textit{DeepFilterNet}& 87.65 & 89.34 & 85.06 & 85.96 &
  \textit{DeepFilterNet}& 62.60 & 53.60 & 46.20 & 38.20 \\
  \Xhline{2pt}
  \end{tabular}%
  }
  \caption{Denoising mitigation for StepAudio2 across two acoustic scenarios.
  The table reports StepAudio2 performance under increasing multi-source acoustic interference ($K=1\ldots4$) in Pasture and Classroom.
  Each task block compares the no-denoise baseline (Noise) with four denoising methods; ASR is WER (lower is better) and other tasks are higher-is-better.}
  \label{tab:mitigation_step2_pasture_classroom}
  \end{table*}

\label{sec:appendix}

\end{document}